\documentclass[aps, prl, reprint, superscriptaddress]{revtex4-2}

\usepackage{ulem}

\usepackage{amsmath}
\usepackage{amssymb}
\usepackage{bm}
\usepackage{braket}
\usepackage[notrig]{physics}
\usepackage{float}
\usepackage{xcolor}
\usepackage{graphicx}
\usepackage[mode=buildnew]{standalone}
\usepackage[caption=false]{subfig}

\usepackage{circuitikz} 
\usepackage[mode=buildnew]{standalone}

\usepackage[colorlinks=true, allcolors=blue]{hyperref}
\usepackage[capitalise]{cleveref}
\AddToHook{cmd/appendix/before}{%
  \crefalias{section}{appendix}%
  \crefalias{subsection}{appendix}%
  \crefalias{subsubsection}{appendix}%
}
\usepackage{orcidlink}

\newcommand{\Z}{\mathbb{Z}}
\newcommand{\Zm}{\bar{Z}_\mathrm{m}}
\newcommand{\Xm}{\bar{X}_\mathrm{m}}
\newcommand{\Ym}{\bar{Y}_\mathrm{m}}
\newcommand{\eps}{\varepsilon}
\newcommand{\sgn}{\mathrm{sgn}}

\newcommand{\inv}{^{-1}}

\newcommand{\ketbasis}[2]{\ket*{#1}_{#2}}
\newcommand{\basisbra}[2]{\,{}_{#1\!\!}\bra*{#2}}

\newcommand{\ketbasisbra}[3]{\ket*{#1}_{#2}\!\!\bra*{#3}}

\renewcommand{\H}{\hat{H}}
\newcommand{\U}{\hat{U}}

\newcommand{\n}{\hat{n}}
\renewcommand{\P}{\hat{P}}
\newcommand{\thetah}{\hat{\theta}}

\newcommand{\varphih}{\hat{\varphi}}
\newcommand{\zetah}{\hat{\zeta}}
\newcommand{\alphah}{\hat{\alpha}}

\newlength{\SingleColLineWidth}
\newlength{\DoubleColLineWidth}
\newcommand\Yale{
    Department of Applied Physics, 
    Yale University, 
    New Haven, CT 06520, USA.
    }
    
\newcommand\YQI{
    Yale Quantum Institute, 
    Yale University, 
    New Haven, CT 06511, USA.
    }
    
\newcommand\USYD{
    School of Physics, 
    University of Sydney 
    Sydney, NSW 2006, Australia.
    }

\let\oldsection\section
\makeatletter
\renewcommand\section[1]{%
	\par\addvspace{1.2ex}%
	\textit{#1---}%
}
\makeatother

\begin{document}
	
	\title{Protected measurements for protected superconducting qubits}
    
	\author{Xanda~C.~Kolesnikow}
	\email{xkol5336@uni.sydney.edu.au}
    \affiliation{\USYD}
    
	\author{Thomas~B.~Smith}
	\affiliation{\Yale}
    \affiliation{\YQI}
    
	\author{Andrew~C.~Doherty}
	\affiliation{\USYD}
    
	\date{\today}
	
	\begin{abstract}
        Protected superconducting qubits such as the $0$-$\pi$ qubit promise to substantially suppress error rates, facilitating fault-tolerant quantum computing with fewer qubits.
        Measuring these qubits is challenging due to their protected nature, and thus far no concrete proposal exists for how to measure them without breaking their protection.
        Here we show how to perform protected measurements of the $0$-$\pi$ qubit in two orthogonal bases.
        The protection of these measurements is facilitated by their quantum non-demolition nature, allowing faults on ancillary measurement qubits to be tolerated.
        As experimental progress pushes protected qubits further into the low error-rate regime, our techniques will be crucial for fault-tolerant universal control.
	\end{abstract}
	
	\maketitle
	
    Fault-tolerant quantum computation with quantum error-correcting codes requires physical qubits to be below a certain threshold error rate~\cite{Kitaev1997,Aharonov1997,Knill1998}.
    Engineering qubits with error rates that are well below this threshold can significantly lower the overhead required to implement quantum algorithms.
	Protected superconducting qubits promise to achieve this
    through the use of circuits whose encoded logic couples weakly to the electromagnetic environment~\cite{Kitaev2006,Gyenis2021a}.
    
	Many theoretical proposals and experimental realizations of protected superconducting qubits have been made recently~\cite{Kalashnikov2020,Smith2020,Larsen2020,Rymarz2021,Gyenis2021,Smith2022,Leroux2023,Ciaccia2024,Guo2024,Kumar2024,Leblanc2025,Smith2025,Lieu2025,Hays2025,Nguyen2025,Kolesnikow2026,Shagalov2026,Guo2026,Messelot2026,Roverch2026,Aramburu2026}.
	However, measuring these qubits remains an outstanding challenge since they are intentionally decoupled from their environment. 
	Traditional methods for superconducting qubit readout rely on non-vanishing qubit matrix elements with linear charge or flux operators.
    These are precisely the matrix elements that are suppressed in a protected superconducting qubit.
	Whilst it is possible to perform unprotected measurements of these qubits by temporarily breaking their protection, doing so will lead to measurement errors becoming rate-limiting and ultimately defeat the purpose of having a protected qubit.
	
	In this work we propose methods for both $Z$-  and $X$-basis measurements of the $0$-$\pi$ qubit~\cite{Kitaev2006,Brooks2013}, a protected superconducting qubit that has been realized in a number of experiments~\cite{Gyenis2021,Kim2024,Hassani2024}.
	We show how to perform these measurements by capacitively coupling the $0$-$\pi$ qubit to ancillary hardware and applying DC flux pulses and AC charge drives.
    We demonstrate that the measurements are protected by showing that their errors decrease exponentially in a circuit or control parameter, and that they are tolerant to imperfect control pulses.
    Such features are a hallmark of fault-tolerant operations on protected qubits~\cite{Brooks2013,Gyenis2021a}.
	Furthermore, our measurements are quantum non-demolition (QND) in nature.
    This permits the use of ancillary qubits that need not be protected --- by resetting and repeating a measurement $N$ times, the overall measurement is tolerant to $(N-1)/2$ errors on the ancillary hardware.

    The protected measurements described in this work, when combined with the single- and two-qubit phase gates proposed in Refs.~\cite{Kitaev2006,Brooks2013,Kolesnikow2026}, unlock a universal, fault-tolerant~\footnote{Our notion of fault-tolerance aligns with the definition of fault-tolerance used for bosonic codes~\cite{Gottesman2001}, whereby the error rate for a fault-tolerant gadget can be exponentially suppressed in a code parameter, e.g., the squeezing of a GKP state. In the protected qubit setting, the code parameter is controlled by a circuit or control parameter, e.g., $\sqrt{E_J/E_C}$.} control scheme for the $0$-$\pi$ qubit.
    Clifford operations are achieved through a combination of single- and two-qubit phase gates, preparation of Pauli eigenstates $\{\ket{0}, \ket{1}, \ket{+}, \ket{-}\}$ via protected measurements, and the Hadamard gate.
    The latter can implemented with gate teleportation as described in Ref.~\cite{Kitaev2006}.
    Non-Clifford operations may be obtained through magic-state injection, or directly implemented using a
    protected non-Clifford gate as proposed in Refs.~\cite{Kolesnikow2026,OBrien2025,Nguyen2025a}.

    A key technical insight of our work is to view protected qubits through the lens of quantum error correction.
    We show that the code structure inherent to a protected qubit may be used to substantially decrease logical errors during measurement.

    \begin{figure}
		\centering
        \subfloat{\label{fig:ideal-Z-meas}}%
        \subfloat{\label{fig:ideal-X-meas}}%
        \includegraphics[width=\SingleColLineWidth]{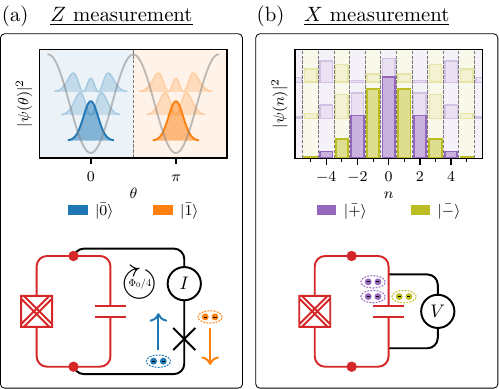}%
		\caption{\textbf{Protected measurements.}
        Shown for an ideal protected qubit described by \cref{eqn:H-ideal-qubit}.
        \textbf{(a)} $Z$ measurements identify anything closer to $\theta = 0$ as a logical $\ket{\bar{0}}$ (blue) and anything closer to $\theta = \pi$ as a logical $\ket{\bar{1}}$ (orange).
        This can be implemented by measuring the direction of the current through a shunting Josephson junction.
        \textbf{(b)} $X$ measurements identify any state with even Cooper-pair parity as a logical $\ket{\bar{+}}$ (purple) and anything with odd Cooper-pair parity as a logical $\ket{\bar{-}}$ (green). 
        This can be implemented by measuring the voltage across the circuit.}
        \label{fig:ideal-qubit-meas}
	\end{figure}%

	\section{Ideal protected qubit}%
	To elucidate the basic principles of protected measurement, we first describe their implementation for an ideal protected superconducting qubit.
    The circuit is depicted in red in \cref{fig:ideal-qubit-meas}, and has the following Hamiltonian:
	\begin{equation}
		\H = 4 E_C \n^2 - E_J \cos2\thetah, \label{eqn:H-ideal-qubit}
	\end{equation}
	where $\thetah$ and $\n$ are the dimensionless flux and charge operators satisfying $[\thetah, \n] = i$. 
	Here we treat $\theta$ as a compact variable residing in the domain $\theta \in [0, 2\pi)$.
    This is the prototypical $\cos{2\theta}$ qubit, upon which many protected qubit proposals are based~\cite{Kitaev2006,Brooks2013,Smith2020}.
	The second term in \cref{eqn:H-ideal-qubit} corresponds to a $4e$-tunneling element, denoted by the double-crossed box in \cref{fig:ideal-qubit-meas}.
    In the protected regime where $E_J \gg E_C$, this Hamiltonian has two nearly degenerate ground states that are localized at the minima of the $\cos2\theta$ potential: $\theta = 0$ and $\theta = \pi$.
	These states are approximate stabilizer states of a rotor-GKP code~\cite{Gottesman2001,Smith2020,Kolesnikow2026}. 
    This code is defined by the stabilizer generators 
    $\hat{S}_{Z_\theta} = e^{2i \thetah}$ and $\hat{S}_{X_\theta} = e^{-2i \pi \n}$. 
    The corresponding logical operators are $\bar{Z}_\theta = e^{i \thetah}$ and $\bar{X}_\theta = e^{-i \pi \n}$.
    Note that the stabilizers are enforced in different ways.
    The $Z$-type stabilizers are enforced energetically: they appear in the potential energy of \cref{eqn:H-ideal-qubit}, and the protected regime is precisely when this term dominates over the kinetic energy.
    The $X$-type stabilizer is enforced exactly due to the periodic boundary conditions on the variable $\theta$; the operator $e^{-2i \pi \n} = \hat I$ for a rotor.

    We can harness the code structure of \cref{eqn:H-ideal-qubit} by judiciously choosing our measurement operators $\Zm$ and $\Xm$.
    These operators define how the logical information is encoded into the physical Hilbert space.
    For example, if we were to construct $\Zm$ by simply projecting onto the two ground states highlighted in \cref{fig:ideal-Z-meas}, then any excitation outside the ground subspace would lead to a logical error, as it would correspond to leakage.
    However, this construction does not account for the correctable errors of the code.
    In the parameter regime of interest, the intrawell transitions dominate over the interwell transitions; the latter are exponentially suppressed in $\sqrt{E_J/E_C}$.
    If we instead choose $\Zm = \sgn[\cos\thetah]$ as our measurement operator then, as shown in \cref{fig:ideal-Z-meas}, any state that is closer to $\theta = 0$ (blue) decodes to $\ket{\bar 0}$ and any state closer to $\theta = \pi$ (orange) decodes to $\ket{\bar 1}$.
    This measurement operator naturally corrects for intrawell transitions.
    In this way, the choice of measurement operator defines a decoding strategy and the choice $\Zm = \sgn[\cos\thetah]$ corresponds to a binning decoder.
    
    The dual measurement operator $\Xm$ can be constructed in a similar fashion.
    As shown in \cref{fig:ideal-X-meas}, states with support on even $n$ (purple) are decoded to $\ket{\bar{+}}$ and states with support on odd $n$ (green) are decoded to $\ket{\bar{-}}$.
    Due to the periodicity of $\theta$, the $X$ measurement operator $\Xm = \sgn[\cos(\pi\n_\theta)]$ coincides with the logical operator $\bar{X}_\theta$.
    This choice of measurement operator corrects for errors that preserve Cooper-pair parity.

    The measurement operators $\Zm$ and $\Xm$ described above correspond to measuring the phase and modular charge across the $\cos{2\theta}$ circuit, which is a direct consequence of the rotor-GKP encoding.
    Ideas for how to measure these operators are given in Refs.~\cite{Kitaev2006a,Kitaev2006,Brooks2013}.
    A measurement of $\Zm$ may be obtained by recording the current through a Josephson junction that forms an inductive loop with the protected qubit.
    When a quarter flux quantum is threaded through the loop, then the direction of the current corresponds to whether the qubit is on a logical $\ket{\bar{0}}$ or $\ket{\bar{1}}$ state.
    Meanwhile, a measurement of $\Xm$ can be achieved by recording the voltage across the circuit.
    Dividing the voltage by $4e$ and binning its value modulo 2 determines whether the qubit is in a logical $\ket{\bar{+}}$ or $\ket{\bar{-}}$ state.
    
    Based on this example, we adopt the following method for designing protected measurements of protected superconducting qubits.
    First, identify the underlying code structure from the circuit Hamiltonian.
    Second, identify a decoding strategy by taking into consideration the most likely errors, and then construct the corresponding measurement operators.
    Third, design ancillary control circuitry to extract information for the chosen measurement operators without compromising the protection of the qubit.
    
    This procedure is made more complex than the simple case of the ideal protected qubit discussed above due to the fact that physically realizable protected superconducting qubits usually involve multiple modes~\cite{Gyenis2021a}. 
	In the remainder of this work, we follow this method to develop protected measurement protocols for the $0$-$\pi$ qubit.

    \begin{figure*}[t]
		\centering
        \subfloat{\label{fig:Z-meas-protocol}}%
        \subfloat{\label{fig:X-meas-protocol}}%
        \includegraphics[width=\DoubleColLineWidth]{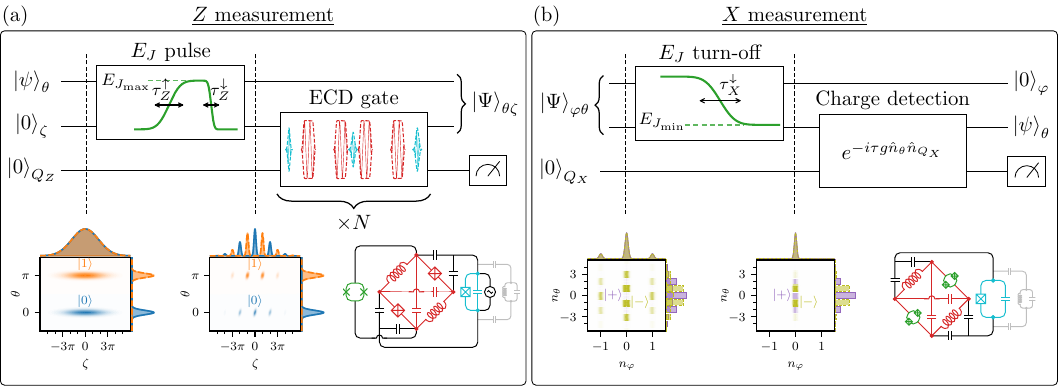}%
		\caption{\textbf{Measuring the $\bm{0}$-$\bm{\pi}$ qubit.} 
            \textbf{(a)} $Z$ measurement. The protocol consists of smoothly turning on the shunting Josephson element (green) to prepare GKP states in the $\zeta$ mode of the $0$-$\pi$ qubit (red) that are entangled with its $\theta$ mode, then sharply turning off the Josephson element, returning the $\zeta$ mode to a linear oscillator.
            Through a series of Echoed Controlled Displacement (ECD) gates, the logical information in the GKP states is transferred to an ancillary qubit (cyan), which is then measured in a single shot via a readout resonator (gray).  
            \textbf{(b)} $X$ measurement. By adiabatically turning off the internal Josephson junctions (green) of the $0$-$\pi$ qubit (red), the $X$-basis information is mapped to the Cooper-pair parity of the $\theta$ mode. 
            The momentum wavefunctions show how the $\ket{\pm}$ states get mapped to states with support on only even or odd $n_\theta$.
            A measurement of a capacitively coupled ancillary qubit (cyan) via a readout resonator (gray) can be used to measure the charge on the $\theta$ mode and infer the $X$-basis information of the $0$-$\pi$ qubit.
            \label{fig:zero-pi-measurement-circuit}
		}
	\end{figure*}
	
	\section{\texorpdfstring{$0$-$\pi$}{0-π} qubit}%
	We start by briefly introducing the $0$-$\pi$ qubit.
	The circuit diagram for this qubit is shown in red in \cref{fig:zero-pi-measurement-circuit}.
	Its Hamiltonian is $\H = \H_{\theta\varphi} + \H_\zeta$, where 
	\begin{align}
		\H_{\theta\varphi} &= 4 E_{C_\theta} \n_\theta^2 + 4 E_{C_\varphi} \n_\varphi^2 + E_L \varphih^2 - 2 E_J \cos \thetah \cos \varphih, \label{eqn:H-theta-phi}\\
		\H_\zeta &= 4 E_{C_\zeta} \n_\zeta^2 + E_L \zetah^2. \label{eqn:H-zeta}
	\end{align}
	Here $[\alphah, \n_\alpha] = i$, with $\alpha = \theta, \varphi, \zeta$ denoting the three modes; $E_{C_\alpha}$ are their charging energies, $E_L$ is the inductive energy for the $\varphi$ and $\zeta$ modes, and $E_J$ is the Josephson energy for the nonlinear coupling between the $\theta$ and $\varphi$ modes.
	The qubit is encoded into the $\theta$ and $\varphi$ modes, whereas $\zeta$ is a linear mode that is decoupled when the circuit elements are symmetric.
	Whilst the $\zeta$ mode plays no role in the qubit encoding, it may be used to perform a protected gate~\cite{Kolesnikow2026}, and here it will be used as an auxiliary mode to perform a $Z$ measurement. 
	Exact expressions for the Hamiltonian parameters and details of the circuit quantization are provided in \cref{app:circuit-quantization}.

    As per our design method, we first identify the underlying code structure.
    The ground states of \cref{eqn:H-theta-phi} are approximate codewords of a concatenated rotor-oscillator-GKP code.
    The stabilizer generators of this code are $\hat S_{Z} = \bar Z_\theta \bar Z_\varphi = e^{i(\thetah + \varphih)}$ (along with $\bar Z_\theta \bar Z_\varphi\inv$) and $\hat S_{X} = \hat S_{X_\theta} = e^{-2i \pi \n_\theta}$.
    As before, $Z$-type stabilizers are enforced at the level of the Hamiltonian in \cref{eqn:H-theta-phi} and $X$-type stabilizers follow from the periodicity of the $\theta$ mode.
    The logical operators for this concatenated code are $\bar Z = \bar Z_\theta = e^{i \thetah}$ and $\bar X = \bar X_\theta \bar X_\varphi = e^{-i \pi (\hat n_\theta + \hat n_\varphi)}$.
    
    Next, we choose our measurement operators.
    We choose a $Z$-type measurement operator that is identical to the one for the ideal protected qubit: $\Zm = \sgn[\cos\thetah]$.
    However, the $X$-type measurement operator takes the form $\Xm = \sgn[\cos(\pi\n_\theta + \pi \n_\varphi)]$, reflecting the fact that the logical qubit is jointly encoded in the $\theta$ and $\varphi$ modes.
    In \cref{app:logical-subsystem} we explain how these measurement operators define a subsystem decomposition of the full Hilbert space.
    For the remainder of this work, we describe the additional circuitry required to measure these operators.
    
	\section{\texorpdfstring{$Z$}{Z} measurement}%
    The natural extension of the $Z$-basis measurement scheme for the ideal protected qubit to the $0$-$\pi$ qubit would be to shunt the $0$-$\pi$ circuit with a Josephson junction.
	However, as pointed out in Refs.~\cite{Paolo2019,Kolesnikow2026} this leads to nonlinear coupling between the $\theta$ and $\zeta$ modes.
	If we assume the $\zeta$ mode is in its ground state, this interaction exponentially suppresses the strength of the current through the shunting Josephson junction in the impedance of the $\zeta$ mode~\cite{Kolesnikow2026}. 
	Since a high impedance of the $\zeta$ mode is required for the $0$-$\pi$ qubit to be protected, this approach is infeasible.
    
    Instead, we propose to measure the $0$-$\pi$ qubit in the $Z$ basis using the protocol shown in \cref{fig:Z-meas-protocol}.
    Here, the $\zeta$ mode is exploited as an auxiliary mode to measure the qubit.
    We modulate the nonlinear interaction (green), $-E_{J_\mathrm{int}}(t) \cos(\thetah + \zetah)$, such that the $\zeta$ mode is prepared in a logical GKP state that is entangled with the state of the $0$-$\pi$ qubit.
    The $Z$-basis information stored in the $\zeta$ mode may then be transferred to an ancillary transmon (cyan), and finally extracted via a readout resonator (gray).
    
    The flux pulse required to entangle the $\zeta$ mode is shown in green in \cref{fig:Z-meas-protocol}, and the wavefunctions before and after this pulse is applied are shown below it.
    The flux is ramped up over a timescale $\tau_Z^\uparrow \sim 1/\omega_\zeta$, where $\omega_\zeta$ is the frequency of the $\zeta$ mode.
    The nonlinear coupling prepares a GKP state in the $\zeta$ mode, which starts in an oscillator ground state.
    This is the same mechanism used for the protected gate in Ref.~\cite{Kolesnikow2026}.
    Once the GKP state has been prepared, the flux is turned off over a much shorter timescale $\tau_Z^\downarrow \ll 1/\omega_\zeta$.
    If this is done sufficiently rapidly, then the $\zeta$ mode returns to the linear Hamiltonian in \cref{eqn:H-zeta}, but remains in a GKP state.
    We note that the measurement is insensitive to when the pulse is turned off; leaving the interaction on amounts to a logical $Z$-rotation, which does not affect the outcome of the $Z$-basis measurement because it commutes with the measurement basis. 

    An ancillary transmon is used to extract the logical information encoded in the $\zeta$ mode.
    The transmon is capacitively coupled to the $0$-$\pi$ qubit as shown in \cref{fig:Z-meas-protocol}.
    A sequence of voltage pulses can be used to perform an Echoed Controlled Displacement (ECD) gate~\cite{Fluhmann2018,Touzard2019,CampagneIbarcq2020}, which entangles the $\zeta$ mode and the ancillary transmon.
    Repeated applications of the ECD gate with different single-qubit rotations can be used to map the logical information onto the ancillary transmon.
    In \cref{app:Z-meas-error-comparison} we discuss the optimal mapping, which can be achieved via Quantum Signal Processing~\cite{Chalermpusitarak2026}.
    Once the information has been mapped to the ancillary transmon, a final measurement of the transmon via a readout resonator facilitates a single-shot measurement in the $Z$ basis.

    Since the ancillary transmon is capacitively coupled to the $\zeta$ mode, the full Hamiltonian commutes with the measurement operator $\Zm = \sgn[\cos\thetah]$.
    This means that this measurement scheme is QND.
    Therefore, by resetting the $\zeta$ mode using the cooling procedure in Refs.~\cite{Paolo2019,Kolesnikow2026} and repeating the measurement protocol, errors on both the $\zeta$ mode and the ancillary transmon may be tolerated.
    It also means that repeated measurements can be used to prepare the $0$-$\pi$ qubit in $\ket{\bar 0}$ or $\ket{\bar 1}$.
    \Cref{app:Z-state-prep} contains more details on fault-tolerant state preparation.

\begin{figure}[t]
        \subfloat{\label{fig:Z-meas-errors}}%
        \subfloat{\label{fig:X-meas-errors}}%
        \centering
        \includegraphics[width=\SingleColLineWidth]{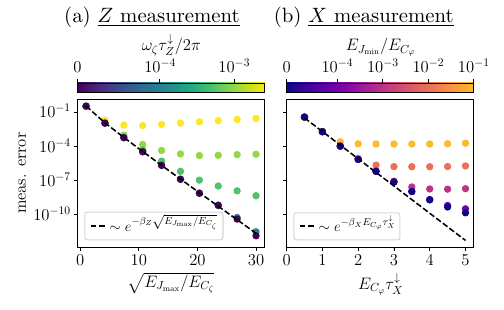}
        \caption{\textbf{Measurement errors.}
        \textbf{(a)} $Z$ measurement errors as a function of the maximum Josephson energy $E_{J_\mathrm{max}}$ for different turn-off times $\tau^\downarrow_Z$.
        The black dashed line is an analytically calculated measurement error; see \cref{app:Z-meas-error-comparison}.
        The $\zeta$-mode energy ratio is $E_L/E_{C_\zeta} = 4.1 \times 10^{-3}$, and $\theta$ is treated as a classical variable.
        \textbf{(b)} $X$ measurement errors as a function of the turn-off time $\tau^\downarrow_X$ for different minimum Josephson energies $E_{J_\mathrm{min}}$.
        The black dashed line is an exponential fit to the $E_{J_\mathrm{min}} = 0$ and $E_{C_\varphi}\tau^\downarrow_X < 3$ data. 
        The $0$-$\pi$ qubit parameters are $E_{C_\theta}/E_{C_\varphi} = 10^{-2}$, $E_J/E_{C_\varphi} = 2.5$ and $E_L/E_{C_\theta} = 4.1 \times 10^{-3}$.
        See \cref{app:numerical-simulations} for details of the numerical simulations.}
    \end{figure}

    In \cref{fig:Z-meas-errors} we plot the measurement error for this protocol and show that it is exponentially suppressed in the maximum strength of the nonlinear coupling $E_{J_\mathrm{max}}$.
    This is due to the fact that a larger $E_{J_\mathrm{max}}$ prepares a GKP state with a higher position-quadrature squeezing in the $\zeta$ mode, which is the quadrature that contains the $Z$-basis information.
    In \cref{app:Z-meas-error-comparison} we provide an analytical estimate for this measurement error [shown by the black line in \cref{fig:Z-meas-errors}], explaining why it scales exponentially with $\sqrt{E_{J_\mathrm{max}}/E_{C_\zeta}}$.
    This exponential scaling is indicative of the protected nature of the measurement.

    Once the GKP state is prepared in the $\zeta$ mode, the nonlinear coupling should be turned off as rapidly as possible.
    This is to prevent distortion of the GKP state due to relaxation.
    In \cref{fig:Z-meas-errors} we find that a slower turn-off disrupts the exponential suppression of the measurement error, and introduces a floor when $\tau^\downarrow_Z$ is larger than $0.1\%$ of a harmonic oscillator period.
    This is quite a fast timescale for typical oscillator frequencies.
    However, the $\zeta$ mode is naturally low frequency in the protected regime of the $0$-$\pi$ qubit.
    Furthermore, here we have used an error function ramp for the turn-off, but it may be possible to optimize the pulse to improve the diabaticity for the same turn-off time.
    In reality, whatever pulse shape results in the lowest measurement error should be chosen, and this will depend on the properties of the classical electronics used to generate the signal.

    For simplicity, the simulations shown here model the dynamics of the $\zeta$ mode only; $\theta$ is treated as a classical binary variable that takes values $0$ or $\pi$.
    This assumption is justified in the protected regime where $E_J \gg E_{C_\theta}$.
    In \cref{app:Z-meas-eff-model} we perform simulations that capture the dynamics of all three modes of the $0$-$\pi$ qubit during the $Z$ measurement to show that the measurement errors are unaffected in the protected regime.

    \section{\texorpdfstring{$X$}{X} measurement}%
    Unlike the ideal protected qubit, a direct measurement of the Cooper-pair parity of the $\theta$ mode does not yield an $X$-basis measurement.
    This is because the logical $X$-basis information is encoded not just in $\theta$, but jointly in the $\theta$ and $\varphi$ modes of the $0$-$\pi$ qubit through the $-2E_J\cos\thetah\cos\varphih$ potential in \cref{eqn:H-theta-phi}.
    
    To obtain an $X$ measurement, this cosine potential must be turned off by tuning the internal Josephson junctions of the $0$-$\pi$ qubit as shown in \cref{fig:X-meas-protocol}.
    This turn-off decouples the $\theta$ and $\varphi$ modes of the $0$-$\pi$ qubit, and leaves the $\theta$ mode in a superposition of only even or odd Cooper-pair number states depending on whether the $0$-$\pi$ qubit was in the logical $\ket{\bar +}$ or $\ket{\bar -}$ state, as shown by the wavefunctions.
    Importantly, while tuning $E_J \to 0$ breaks the protection of the $0$-$\pi$ qubit, it does so only in the $Z$ basis because the charge islands in the $0$-$\pi$ qubit are preserved. 

    Measuring the Cooper-pair parity on the $\theta$ mode then enacts an $X$-basis measurement.
	To measure this charge, the $\theta$ mode of the $0$-$\pi$ qubit may be capacitively coupled to an ancillary unprotected qubit as shown in the circuit diagram in \cref{fig:X-meas-protocol}.
	This leads to a shift of the ancillary qubit frequency depending on the charge residing on the $\theta$ mode, which can be measured via a capacitively coupled readout resonator.
	It is important to note that this dispersive shift is due to a DC offset charge.
	This means that the ancillary qubit should not be too far into the transmon regime so that it remains sensitive to changes in DC charge.
	In practice, a balance should be struck to minimize the errors in the ancillary qubit that occur due to increased charge sensitivity to the environment and longer wait-times from a small dispersive shift.
	
	Once again, an important feature of this measurement scheme is that it is exactly QND, assuming $E_J$ can be brought to zero.
	This is because when $E_J = 0$ in \cref{eqn:H-theta-phi} (and the $0$-$\pi$ qubit is charge-coupled to an ancillary qubit) the operator $e^{-i \pi \n_\theta}$ commutes with the system Hamiltonian.
	Therefore, the ancillary qubit may be reset and remeasured multiple times to tolerate errors on the noisier ancillary qubit.
    Furthermore, to prepare the $0$-$\pi$ qubit in the $\ket{\bar+}$ (or $\ket{\bar-}$) state, the measurement procedure can be repeated until a measurement of $+1$ (or $-1$) outcome is achieved.
    Then, by reversing the flux pulse to tune $E_J$ from 0, a $\ket{\bar+}$ (or $\ket{\bar-}$) state will be prepared in the $0$-$\pi$ qubit.
    For more details on state preparation see \cref{app:X-state-prep}.
    
	In \cref{fig:X-meas-errors} we plot the measurement error as a function of the ramp turn-off time $\tau_X^\downarrow$. 
    Here we observe that the measurement error initially decreases exponentially with a decay constant on the order of the $\varphi$-mode charging energy.
    This is because parity-changing transitions of the $\theta$ mode are heavily suppressed in the small $E_L$ regime.
    See \cref{app:adiabatic-turn-off} for a more detailed discussion of the adiabaticity requirements.
    Given that the $\varphi$ mode of the $0$-$\pi$ qubit naturally has a large charging energy, a turn-off time on the order of $1/E_{C_\varphi}$ should be well within the relaxation time of the ancillary qubit.
    The departure from the exponential decay for larger values of $\tau_X^\downarrow$ is due to the non-zero value of $E_L$.
    In \cref{app:impedance} we study how the measurement error changes with $E_L$.
    
    In practice, the tunable Josephson junctions cannot be switched off completely.
    This leads to imperfect charge states as well as non-QNDness.
    In \cref{fig:X-meas-errors} we also plot the measurement error as a function of the turn-off time with non-zero final values of $E_J$.
    Here, we find that so long as $E_J$ is tuned to within $1\%$ of its maximum value, measurement errors below $10^{-5}$ may still be reached with the same measurement time. 
    To reach lower error rates, tunable Josephson elements made from multiple flux loops can be used to reach smaller $E_J$ values than are permitted from asymmetries in Josephson energies in a SQUID~\cite{Beauseigneur2026,Kolesnikow2026}.
	
	\section{Conclusion and outlook}%
    In this work we have explained how to perform protected measurements of protected superconducting qubits.
	In particular, we have shown how to perform QND measurements of the $0$-$\pi$ qubit in the $Z$ and $X$ bases such that the ancillary hardware does not compromise the protection of the qubit. 
    Our protocols were obtained by considering the underlying code structure of the $0$-$\pi$ qubit.
	Together with the single- and two-qubit protected phase gates proposed in Ref.~\cite{Kolesnikow2026} this completes the set of tools required for universal fault-tolerant operations with the $0$-$\pi$ qubit~\cite{Kitaev2006,Brooks2013}.
	We envisage a fully protected qubit implementation based on these building blocks.
	
    While we have focused on the $0$-$\pi$ qubit here, our code-inspired procedure should apply to most protected superconducting qubits.
    Furthermore, numerous protected superconducting qubits possess a GKP-like encoding~\cite{Le2019,Smith2020,Rymarz2021,Vuillot2024,Nguyen2025}.
    The current-charge duality elucidated in our example of the ideal protected qubit serves as useful intuition for measuring these qubits.
    As highlighted here, the multiple modes of these circuits must be carefully considered when designing a measurement scheme, and thus the specific implementation will strongly depend on the circuit for the protected qubit.
    Designing such protocols represents an important area for future work as experimental progress on these qubits matures.
    
    If one wishes to make use of the substantially suppressed error rates promised by protected qubits, protected measurements of the kind described in this work are critical.
    Without fault-tolerant operations, the use case of a protected qubit is strongly diminished, as it is only as good as its most erroneous operation.

    \section{Acknowledgments}%
    We thank Xanthe Croot for discussions about experimental constraints, and Teerawat Chalermpusitarak and Ting Rei Tan for discussions about measuring bosonic observables via bosonic Quantum Signal Processing.
    X.C.K. was supported by an Australian Government Research Training Program (RTP) Scholarship and acknowledges computational resources from Australia's National Computational Infrastructure (NCI), via the University of Sydney's Access Scheme. 
    X.C.K. also acknowledges support from the Intelligence Advanced Research Projects Activity (IARPA) under the Entangled Logical Qubits program through Cooperative Agreement Number W911NF-23-2-0223. 
    The views and conclusions contained in this document are those of the authors and should not be interpreted as representing the official policies, either expressed or implied, of IARPA, the Army Research Office, or the U.S. Government. 
    The U.S. Government is authorized to reproduce and distribute reprints for Government purposes notwithstanding any copyright notation herein.
    We acknowledge the traditional owners of the land on which this work was undertaken at the University of Sydney, the Gadigal people of the Eora Nation.
    
    Python code for this manuscript was produced with assistance from Claude Opus 4.8 and ChatGPT 5.5. Claude Fable 5 gave us the idea to use a WKB calculation to estimate the error rate in \cref{app:Z-meas-error-comparison}. 
    All code output and calculations have been verified by the authors.
    Manuscript writing was done without AI tool assistance.

    \section{Data availability}%
    The data that support the findings of this article are openly available~\cite{ZenodoRepo}.

    \newpage
	\appendix
	\onecolumngrid
	
	\let\section\oldsection
    \setcounter{secnumdepth}{3}

    \section{Circuit quantization} \label{app:circuit-quantization}
    
    In this section, we outline our circuit quantization procedure.
    This procedure is described in more detail in Appendix A of Ref.~\cite{Kolesnikow2026}, which is based on the work of Refs.~\cite{Rajabzadeh2023,Vool2017}.
    We consider the full circuit shown in \cref{fig:zero-pi-measurement-circuit-app}, which includes all ancillary components necessary for measuring the $0$-$\pi$ qubit in both bases.
    We omit the readout resonators for simplicity, as their coupling to the ancilla qubits (cyan) simply leads to a renormalization of the charging energies for the circuit components considered here.
    
    We quantize the circuit in \cref{app:circuit-methodology}, and present the resulting Hamiltonian in \cref{app:circuit-Hamiltonian}. 
    In \cref{app:Z-Hamiltonian,app:X-Hamiltonian}, we show how this Hamiltonian is used for measurement in the logical $Z$ and $X$ bases.
    In \cref{app:capacitive-disorder} we discuss the effects of capacitive disorder on measurement and qubit coherence.

    \subsection{Methodology} \label{app:circuit-methodology}
    
\begin{figure}
		\centering
        \includegraphics[scale=1.2]{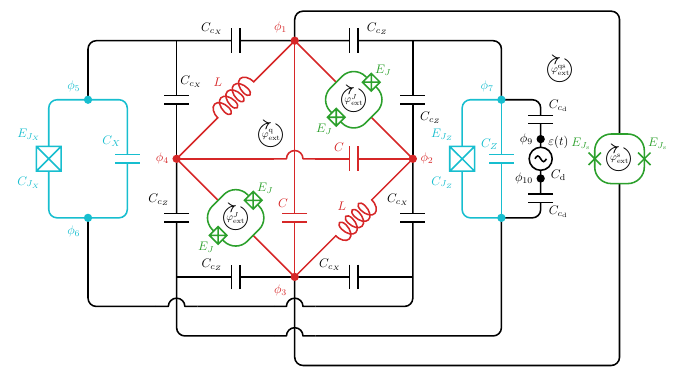}%
		\caption{Complete circuit diagram for measuring the $0$-$\pi$ qubit in both bases.}
        \label{fig:zero-pi-measurement-circuit-app}
	\end{figure}

    The nodes labeled in \cref{fig:zero-pi-measurement-circuit-app} define a graph whose edges correspond to circuit elements.
    The matrix that relates each edge of the circuit to a pair of nodes is
    \begin{equation} 
        \label{eqn:adjacency-matrix}
        \mathbf{A} = 
        \begin{pmatrix}
            -1 & 1 & 0 & 0 & 0 & 0 & 0 & 0 & 0 & 0 \\
            -1 & 1 & 0 & 0 & 0 & 0 & 0 & 0 & 0 & 0 \\
            0 & -1 & 1 & 0 & 0 & 0 & 0 & 0 & 0 & 0 \\
            0 & 0 & -1 & 1 & 0 & 0 & 0 & 0 & 0 & 0 \\
            0 & 0 & -1 & 1 & 0 & 0 & 0 & 0 & 0 & 0 \\
            1 & 0 & 0 & -1 & 0 & 0 & 0 & 0 & 0 & 0 \\
            -1 & 0 & 1 & 0 & 0 & 0 & 0 & 0 & 0 & 0 \\
            0 & 1 & 0 & -1 & 0 & 0 & 0 & 0 & 0 & 0 \\
            -1 & 0 & 1 & 0 & 0 & 0 & 0 & 0 & 0 & 0 \\
            -1 & 0 & 1 & 0 & 0 & 0 & 0 & 0 & 0 & 0 \\
            0 & 0 & 0 & 0 & 1 & -1 & 0 & 0 & 0 & 0 \\
            0 & 0 & 0 & 0 & 1 & -1 & 0 & 0 & 0 & 0 \\
            1 & 0 & 0 & 0 & -1 & 0 & 0 & 0 & 0 & 0 \\
            0 & 0 & 0 & 1 & -1 & 0 & 0 & 0 & 0 & 0 \\
            0 & 1 & 0 & 0 & 0 & -1 & 0 & 0 & 0 & 0 \\
            0 & 0 & 1 & 0 & 0 & -1 & 0 & 0 & 0 & 0 \\
            0 & 0 & 0 & 0 & 0 & 0 & 1 & -1 & 0 & 0 \\
            0 & 0 & 0 & 0 & 0 & 0 & 1 & -1 & 0 & 0 \\
            1 & 0 & 0 & 0 & 0 & 0 & -1 & 0 & 0 & 0 \\
            0 & 1 & 0 & 0 & 0 & 0 & -1 & 0 & 0 & 0 \\
            0 & 0 & 1 & 0 & 0 & 0 & 0 & -1 & 0 & 0 \\
            0 & 0 & 0 & 1 & 0 & 0 & 0 & -1 & 0 & 0 \\
            0 & 0 & 0 & 0 & 0 & 0 & 0 & 0 & 1 & -1 \\
            0 & 0 & 0 & 0 & 0 & 0 & 1 & 0 & -1 & 0 \\
            0 & 0 & 0 & 0 & 0 & 0 & 0 & 1 & 0 & -1
        \end{pmatrix}
        .
    \end{equation}
    Edges of the circuit graph are split into three categories: capacitors, inductors, and Josephson junctions.
    Edges with capacitances define a diagonal capacitance matrix, where the $j$-th element of the diagonal is the capacitance of the $j$-th edge in the circuit graph:
    \begin{equation} 
        \label{eqn:capacitance-matrix}
        \mathbf{C} 
        = 
        \mathrm{diag}(C_J, C_J, 0, C_J, C_J, 0, C, C, 0, 0, C_{J_X}, C_X, C_{c_X}, C_{c_X}, C_{c_X}, C_{c_X}, C_{J_Z}, C_Z, C_{c_Z}, C_{c_Z}, C_{c_Z}, C_{c_Z}, C_\mathrm{d}, C_{c_\mathrm{d}}, C_{c_\mathrm{d}}).
    \end{equation}
    Inductive edges are characterized by a diagonal inverse inductance matrix:
    \begin{equation} 
        \label{eqn:inductance-matrix}
        \mathbf{L}^{-1}
        = 
        \mathrm{diag}(0, 0, L^{-1}, 0, 0, L^{-1}, \dots)
        ,
    \end{equation}
    where the remaining elements are zero.
    Finally, the Josephson junction edges are described by the vector
    \begin{equation}
        \label{eqn:Josephson-energies}
        \mathbf{E}_J 
        = 
        (E_J, E_J, 0, E_J, E_J, 0, 0, 0, E_{J_\mathrm{s}}, E_{J_\mathrm{s}}, E_{J_X}, 0, 0, 0, 0, 0, E_{J_Z}, \dots),
    \end{equation}
    where again the remaining elements are zero.
    The classical Lagrangian for this circuit is $\mathcal L = T - V$, where
    \begin{align}
        &T (\dot{\mathbf \Phi})
        = 
        \frac12 
        \dot{\mathbf \Phi}^\mathrm{T}
        \mathbf C
        \dot{\mathbf \Phi}
        ,
        \\
        &V (\mathbf \Phi)
        =
        \frac12 
        \mathbf \Phi^\mathrm{T}
        \mathbf L\inv
        \mathbf \Phi
        -
        \sum_{j} \left( \mathbf{E}_{J} \right)_j \cos(\mathbf{\Phi}_j/\phi_0)
        ,
    \end{align}
    are the kinetic and potential energy terms, written in terms of the branch flux variables $\mathbf{\Phi}$, and $\phi_0 = \hbar/2e$ is the reduced flux quantum.
    
    In the absence of external magnetic fluxes, the matrix $\mathbf{A}$ defines the correspondence between branch flux variables $\bm{\Phi}$ and node flux variables $\bm{\phi}$.
    In the presence of external magnetic fluxes $\bm{\phi}_\mathrm{ext}$,
    \begin{equation}
        \bm{\Phi} = \mathbf{A}\bm{\phi} + \mathbf{B}\bm{\phi}_\mathrm{ext},
    \end{equation}
    where $\bm{\phi}_\mathrm{ext} = (\phi_\mathrm{ext}^\mathrm{q}, \phi_\mathrm{ext}^J, \phi_\mathrm{ext}^J, \phi_\mathrm{ext}^\mathrm{s}, \phi_\mathrm{ext}^\mathrm{qs})^\mathrm{T}$ are the external magnetic fluxes threading the five inductive loops of the circuit as shown in \cref{fig:zero-pi-measurement-circuit-app}.
    Different choices for $\mathbf{B}$ correspond to different choices of gauge, which affect the form of the Hamiltonian but not the underlying physics.
    The irrotational gauge corresponds to the choice~\cite{You2019}
    \begin{equation}
        \mathbf{B} = 
        \begin{pmatrix}
            \mathbf{A}^\mathrm{T} \mathbf{C} \\
            \mathbf{G}
        \end{pmatrix}^+
        \begin{pmatrix}
            \mathbf{0} \\
            \mathbf{I}
        \end{pmatrix},
    \end{equation}
    where the superscript $+$ denotes the pseudoinverse of the matrix.
    Here $\mathbf{0}$ is a $10 \times 5$ matrix of zeros, $\mathbf{I}$ is the $5 \times 5$ identity matrix and $\mathbf{G}$ is the following matrix: 
    \begin{equation} \label{eqn:flux-loops-matrix}
        \mathbf{G} = 
        \begin{pmatrix}
            1 & 0 & 1 & 0 & 1 & 1 & 0 & 0 & 0 & 0 & 0 & \cdots\\
            -1 & 1 & 0 & 0 & 0 & 0 & 0 & 0 & 0 & 0 & 0 & \cdots\\
            0 & 0 & 0 & 1 & -1 & 0 & 0 & 0 & 0 & 0 & 0 & \cdots\\
            0 & 0 & 0 & 0 & 0 & 0 & 0 & 0 & -1 & 1 & 0 & \cdots\\
            0 & -1 & -1 & 0 & 0 & 0 & 0 & 0 & 1 & 0 & 0 & \cdots
        \end{pmatrix},
    \end{equation}
    where the omitted entries are all zero. 
    This matrix is defined such that $\mathbf{G} \mathbf{\Phi} = \bm{\phi}_\mathrm{ext}$; the sum of branch fluxes around an inductive loop equals the external flux threading that loop.
    In this gauge,
    \begin{align}
        &T(\dot{\bm{\phi}}) 
        = 
        \frac{1}{2} 
        \dot{\bm{\phi}}^\mathrm{T} \mathbf{C}_\phi \dot{\bm{\phi}}
        , 
        \\
        &V(\bm{\phi}) 
        = 
        \frac{1}{2} 
        \dot{\bf{\phi}}^\mathrm{T} \mathbf{L}_\phi^{-1}\bm{\phi} 
        + 
        \bm{\phi}_\mathrm{ext}^\mathrm{T} \mathbf{B}^\mathrm{T} \mathbf{L}^{-1} \mathbf{A} \bm{\phi} 
        - 
        \sum_{j} \left( \mathbf{E}_{J} \right)_j  \cos[(\mathbf{A}_j \bm{\phi} - \mathbf{B}_j \bm{\phi}_\mathrm{ext})/\phi_0]
        ,
    \end{align}
    where $\mathbf{C}_\phi = \mathbf{A}^\mathrm{T} \mathbf{C} \mathbf{A}$ and $\mathbf{L}^{-1}_\phi = \mathbf{A}^\mathrm{T} \mathbf{L}^{-1} \mathbf{A}$ are the capacitance and inverse inductance matrices in the basis of node fluxes, and $\mathbf{A}_j$ and $\mathbf{B}_j$ denote the $j$-th rows of $\mathbf{A}$ and $\mathbf{B}$.
    Importantly, the time-derivative of the external fluxes does not appear in the kinetic energy due to the irrotational gauge choice.

    To separate out the dynamical from the non-dynamical variables, we make a coordinate transformation
    \begin{equation}
        \bm{\theta} = \frac{1}{\phi_0} \mathbf{M}^{-1} \bm{\phi},
    \end{equation}
    where $\bm{\theta} = (\theta, \varphi, \zeta, \theta_X, \theta_Z, \theta_\mathrm{d}, \theta_{g_1}, \theta_{g_2}, \theta_{g_3}, \Sigma)^\mathrm{T}$ are the transformed coordinates and the transformation matrix is
    \begin{equation} \label{eqn:coordinate-transformation-matrix}
        \mathbf{M}^{-1} = \begin{pmatrix}
                \frac{1}{2} & -\frac{1}{2} & -\frac{1}{2} & \frac{1}{2}
                & 0 & 0 & 0 & 0 & 0 & 0
                \\
                \frac{1}{2} & -\frac{1}{2} & \frac{1}{2} & -\frac{1}{2}
                & 0 & 0 & 0 & 0 & 0 & 0
                \\
                \frac{1}{2} & \frac{1}{2} & -\frac{1}{2} & -\frac{1}{2}
                & 0 & 0 & 0 & 0 & 0 & 0
                \\
                0 & 0 & 0 & 0
                & 1 & -1 & 0 & 0 & 0 & 0
                \\
                0 & 0 & 0 & 0
                & 0 & 0 & 1 & -1 & 0 & 0
                \\
                0 & 0 & 0 & 0
                & 0 & 0 & 0 & 0 & 1 & -1
                \\
                -\frac{1}{2\sqrt{3}} & -\frac{1}{2\sqrt{3}}
                & -\frac{1}{2\sqrt{3}} & -\frac{1}{2\sqrt{3}}
                & 0 & 0 & 0 & 0
                & \frac{1}{\sqrt{3}} & \frac{1}{\sqrt{3}}
                \\
                -\frac{1}{2\sqrt{6}} & -\frac{1}{2\sqrt{6}}
                & -\frac{1}{2\sqrt{6}} & -\frac{1}{2\sqrt{6}}
                & 0 & 0
                & \frac{1}{2}\sqrt{\frac{3}{2}}
                & \frac{1}{2}\sqrt{\frac{3}{2}}
                & -\frac{1}{2\sqrt{6}} & -\frac{1}{2\sqrt{6}}
                \\
                -\frac{1}{2\sqrt{10}} & -\frac{1}{2\sqrt{10}}
                & -\frac{1}{2\sqrt{10}} & -\frac{1}{2\sqrt{10}}
                & \sqrt{\frac{2}{5}} & \sqrt{\frac{2}{5}}
                & -\frac{1}{2\sqrt{10}} & -\frac{1}{2\sqrt{10}}
                & -\frac{1}{2\sqrt{10}} & -\frac{1}{2\sqrt{10}}
                \\
                \frac{1}{\sqrt{10}} & \frac{1}{\sqrt{10}}
                & \frac{1}{\sqrt{10}} & \frac{1}{\sqrt{10}}
                & \frac{1}{\sqrt{10}} & \frac{1}{\sqrt{10}}
                & \frac{1}{\sqrt{10}} & \frac{1}{\sqrt{10}}
                & \frac{1}{\sqrt{10}} & \frac{1}{\sqrt{10}}
            \end{pmatrix}
            .
    \end{equation}
    The first five modes are the dynamical variables of interest: $\theta$, $\varphi$ and $\zeta$ are the internal dynamical modes of the $0$-$\pi$ qubit, and $\theta_X$ and $\theta_Z$ are the modes for the ancillary qubits.
    The last five modes are non-dynamical and will not appear as quantum variables in the final Hamiltonian: $\theta_\mathrm{d}$ is the degree of freedom for the charge drive, $\theta_{g_1}$, $\theta_{g_2}$ and $\theta_{g_3}$ are free variables that only appear in the kinetic energy, and $\Sigma$ is a zero-mode, which does not appear in either the kinetic or potential energy and may therefore be discarded. 

    In the transformed coordinates, the kinetic and potential energies become
    \begin{align}
        &T(\dot{\bm{\theta}}) 
        = 
        \frac{\phi_0^2}{2} \dot{\bm{\theta}}^\mathrm{T} \mathbf{C_\theta} \dot{\bm{\theta}}
        , 
        \\
        &V(\bm{\theta},\bm{\varphi}_\mathrm{ext}) 
        = 
        \frac{\phi_0^2}{2} \bm{\theta}^\mathrm{T} \mathbf{L}_\theta^{-1} \bm{\theta} 
        + 
        \phi_0^2 \bm{\varphi}_\mathrm{ext}^T \mathbf{B}^\mathrm{T} \mathbf{L}^{-1} \mathbf{A} \mathbf{M} \bm{\theta}
        ,
    \end{align}
    where $\mathbf{C_\theta} = \mathbf{M}^\mathrm{T} \mathbf{C}_\mathrm{\phi} \mathbf{M}$ and $\mathbf{L}_\theta^{-1} = \mathbf{M}^\mathrm{T} \mathbf{L}^{-1}_\mathrm{\phi} \mathbf{M}$ are the capacitance and inverse inductive matrices in the transformed coordinate basis, and $\bm{\varphi}_\mathrm{ext} = \bm{\phi}_\mathrm{ext}/\phi_0$ are the dimensionless fluxes.
    To eliminate the free variables from the Hamiltonian, we employ the Schur complement method~\cite{Weissler2026} to rewrite the kinetic energy as
    \begin{equation}
        T =  \frac{\phi_0^2}{2} \dot{\bm{\theta}}^\mathrm{T}(\mathbf{C_\theta}/\mathbf{C}_{22})^{-1} \dot{\bm{\theta}} , 
    \end{equation}
    where the reduced capacitance matrix is obtained via the Schur complement:
    \begin{equation}
       \mathbf{C_\theta}/\mathbf{C}_{22} 
       = 
       \mathbf{C}_{11} - \mathbf{C}_{12} \mathbf{C}_{22}^{-1} \mathbf{C}_{21}
       .
    \end{equation}
    Here we have expressed the original capacitance matrix as a block matrix
    \begin{equation}
        \mathbf{C_\theta} = 
        \begin{pmatrix} 
            \mathbf{C}_{11} & \mathbf{C}_{12} \\ 
            \mathbf{C}_{21} & \mathbf{C}_{22}
        \end{pmatrix},
    \end{equation}
    where the indices 1 and 2 label the dynamical and non-dyanmical sectors, respectively.

    Performing a Legendre transformation then yields the classical Hamiltonian $H = T + V$
    where
    \begin{equation}
        T(\mathbf{n}) = \frac{4 e^2}{2} \mathbf{n}^\mathrm{T} \mathbf{C}^{-1}_\theta \mathbf{n},
    \end{equation}
    and $\mathbf{n} = \phi_0 \mathbf{C}_\theta \dot{\bm{\theta}}/2e$
   are the dimensionless charge variables that are canonically conjugate to $\bm{\theta}$.
   Finally, we promote the classical variables to quantum operators $\{ \bm{\theta}, \mathbf{n} \} \to \{ \hat{\bm{\theta}}, \hat{\mathbf{n}} \}$, which satisfy the commutation relations
   \begin{equation}
       \big[ \hat{\bm{\theta}}_j, \hat{\mathbf{n}}_{j'} \big] = i \delta_{j,j'}.
   \end{equation}
    
    \subsection{Hamiltonian} \label{app:circuit-Hamiltonian}
    Setting the flux in the loop between the $0$-$\pi$ qubit and the SQUID shunt to be $\varphi_\mathrm{ext}^\mathrm{qs} = (\varphi_\mathrm{ext}^\mathrm{q} + \varphi_\mathrm{ext}^J + \varphi_\mathrm{ext}^\mathrm{s})/2$, we obtain the following Hamiltonian for the circuit in \cref{fig:zero-pi-measurement-circuit-app}:
    \begin{equation}
        \H = \H_{\theta\varphi} + \H_\zeta + \H_X +\H_Z  + g_X \n_\theta \n_X + g_Z \n_\zeta \n_Z,
    \end{equation}
    where we decompose the full Hamiltonian into the following terms:
    \begin{align}
        &\H_{\theta\varphi} 
        = 
        4 E_{C_\theta} \n_\theta^2 
        + 
        4 E_{C_\varphi} \n_\varphi^2 
        + 
        E_L 
        \left[
            \varphih 
            - 
            \tfrac12
            \left(
            \varphi_\mathrm{ext}^\mathrm{q} + \varphi_\mathrm{ext}^J
            \right)
        \right]^2 
        - 
        2E_J(t) \cos\thetah\cos\varphih
        , 
        \\
        &\H_{\zeta}
        =
        4 E_{C_\zeta} \n_\zeta^2 
        +
        E_L\zetah^2
        - 
        E_{J_\mathrm{int}}(t) \cos(\thetah + \zetah) 
        + 
        \Omega_\zeta(t)\n_\zeta
        \\
        &\H_X 
        = 
        4 E_{C_X} \n_X^2 
        - 
        E_{J_X} \cos\thetah_X
        , 
        \\
        &\H_Z 
        = 
        4 E_{C_Z} \n_Z^2 
        - 
        E_{J_Z} \cos\thetah_Z 
        + 
        \Omega_Z(t) \n_Z
        .
    \end{align} 
    The charging energies are $E_{C_\theta} = e^2/2\tilde C_\theta$, $E_{C_\varphi} = e^2/2\tilde C_\varphi$, $E_{C_\zeta} = e^2/2\tilde C_\zeta$, $E_{C_X} = e^2/2\tilde C_X$ and $E_{C_Z}=e^2/2\tilde C_Z$.
    The inductive energy of the $0$-$\pi$ qubit is $E_L = \phi_0^2/L$.
    The Josephson energies are $E_J(t) = 2 E_J \cos[\varphi_\mathrm{ext}^J(t)/2]$, $E_{J_\mathrm{int}}(t) = 2E_{J_\mathrm{s}}\cos[\varphi_\mathrm{ext}^\mathrm{s}(t)/2]$, $E_{J_X}$ and $E_{J_Z}$. 
    The charge drives are $\Omega_\zeta(t)$ and $\Omega_Z(t)$.
    The capacitive coupling strengths are $g_X$ and $g_Z$.
    For convenience, we often define the $0$-$\pi$ qubit parameter regime in terms of its $\zeta$- or $\varphi$-mode impedance, which are defined by $Z_\zeta = \sqrt{L/4C}$ and $Z_\varphi = \sqrt{L/4C_J}$.
    We quote these values in units of the superconducting resistance quantum $R_Q = h/4e^2$.
    
    The dressed capacitances may be expressed in terms of a difference $\delta C_i$ to their bare capacitances:
    \begin{align}
        \tilde C_\theta &= 2C + 4C_J + \delta C_\theta, \\
        \tilde C_\varphi &= 4 C_J + \delta C_\varphi, \\
        \tilde C_\zeta &= 2C + \delta C_\zeta, \\
        \tilde C_X &= C_X + C_{J_X} + \delta C_X, \\
        \tilde C_Z &= C_Z + C_{J_Z} + \delta C_Z. 
    \end{align}
    The differences are given by the following expressions:
    \begin{align}
        \delta C_\theta &= C_{c_Z} + \frac{C_{c_X} (C_X + C_{J_X})}{C_{c_X} + C_X + C_{J_X}}, \\
        \delta C_\varphi &= C_{c_X} + C_{c_Z}, \\
        \delta C_\zeta &= C_{c_X} + \frac{C_{c_Z}C_{Z\mathrm{d}}}{C_{c_Z}+ C_{Z\mathrm{d}}},  \\
        \delta C_X &= \frac{C_{c_X}(2C + 4C_J + C_{c_Z})}{C_{c_X} + 2C + 4C_J + C_{c_Z}}, \\
        \delta C_Z &= \frac{C_{c_Z}(2C + C_{c_X})}{C_{c_Z} + 2C + C_{c_X}} + \frac{C_{c_\mathrm{d}}  C_\mathrm{d}}{2 C_{c_\mathrm{d}} + 2C_\mathrm{d}},
    \end{align}
    where $ C_{Z\mathrm{d}} = C_Z + C_{J_Z} + {C_{c_\mathrm{d}} C_\mathrm{d}}/({2C_\mathrm{d} + C_{c_\mathrm{d}}})$.
    Note that these all vanish in the limit $C_{c_Z}, C_{c_X}, C_{c_\mathrm{d}} \rightarrow 0$.
    Finally, the capacitive coupling strengths and charge drives are given by
    \begin{align}
        g_X &= \frac{C_{c_X}}{C_{c_X} + C_X + C_{J_X}} 4E_{C_\theta}, \label{eqn:gX}\\
        g_Z &= \frac{C_{c_Z}}{C_{c_Z} + C_{Z\mathrm{d}}} 4E_{C_\zeta} , \\
        \Omega_\zeta(t) &= \frac{C_{c_\mathrm{d}}}{2C_\mathrm{d} + C_{c_\mathrm{d}}} \frac{C_{c_Z}}{C_{c_Z} + C_{Z\mathrm{d}}} 4E_{C_\zeta} \eps(t), \\
        \Omega_Z(t) &= \frac{C_{c_\mathrm{d}}}{2C_\mathrm{d} + C_{c_\mathrm{d}}} 4E_{C_Z} \eps(t),
    \end{align}
    where $\eps(t)$ is the dimensionless drive amplitude.\\

    \subsection{\texorpdfstring{$Z$}{Z}-measurement Hamiltonian} \label{app:Z-Hamiltonian}
    For the $Z$ measurement, $E_J(t) = 2E_J \gg E_{C_\theta}$.
    The effect of the ancillary qubit for the $X$ measurement on the $\theta$ mode becomes exponentially suppressed in $E_J/E_{C_\theta}$.
    In contrast, the interaction between the $\zeta$ mode of the $0$-$\pi$ qubit and the $Z$ measurement ancillary transmon is utilized by driving at the $\zeta$ mode's harmonic oscillator frequency.
    After the GKP states in the $\zeta$ mode have been prepared by tuning $E_{J_\mathrm{int}}(t)$, the coupling between the $\theta$ and $\zeta$ modes is turned off, and so the $\zeta$-$\theta_Z$ subsystem reduces to 
    \begin{equation}
        \H = 4 E_{C_\zeta} \n_\zeta^2 + E_L \zetah^2 + \Omega_\zeta(t) \n_\zeta + 4 E_{C_Z} \n_Z^2 - E_{J_Z} \cos\thetah_Z  + g_Z \n_\zeta \n_Z + \Omega_Z(t) \n_Z,
    \end{equation}
    which corresponds to an oscillator coupled to a qubit with a charge drive. 
    This reduces to the driven Jaynes-Cummings Hamiltonian, which can be used to generate a conditional displacement gate through appropriate drives on the qubit and the $\zeta$ mode~\cite{Touzard2019,CampagneIbarcq2020}.

    \subsection{\texorpdfstring{$X$}{X}-measurement Hamiltonian} \label{app:X-Hamiltonian}
    For the $X$ measurement, $\Omega_\zeta(t) = \Omega_Z(t) = E_{J_\mathrm{int}}(t)= E_J(t) = 0$.
    Thus, all modes in the $0$-$\pi$ qubit decouple from one another.
    The Hamiltonian for the coupled $\theta$ mode and $X$-ancillary qubit is
    \begin{equation}
        \H = 4 E_{C_\theta} \n_\theta^2 + 4E_{C_X} \n_X^2 - E_{J_X}\cos\thetah_X + g_X \n_\theta\n_X,
    \end{equation}
    which may be rewritten as
    \begin{equation}
        \H = \left( 4 E_{C_\theta} - \frac{g_X}{16 E_{C_X}}\right) \n_\theta^2 + 4 E_{C_X}\left(\n_X + \frac{g_X}{8 E_{C_X}} \n_\theta\right)^2 - E_{J_X}\cos\thetah_X.
    \end{equation}
    Demoting $\n_\theta$ to a classical variable since its conjugate flux variable does not appear in the potential energy, the Hamiltonian for the $X$-ancillary qubit corresponds to a transmon or Cooper-pair box with an offset charge of $n_g = g_Xn_\theta/8 E_{C_X}$.
    A measurement of the frequency of the ancillary qubit will reveal the value of the offset charge and therefore the value of $n_\theta$.
    However, the size of this dispersive shift $\chi$ could be quite small in practice if the ancillary transmon is far into the transmon regime ($E_{J_X} \gg E_{C_X}$), leading to a long measurement time $t_m\sim 1/\chi$.
    If instead the ancillary qubit were in the Coper-pair box regime $E_{J_X} \ll E_{C_X}$ then $\chi \approx g_X \le 4E_{C_\theta}$ [\cref{eqn:gX}], and so the measurement time will be limited instead by the charging energy of the $\theta$ mode.
    Whilst a smaller value of $E_{J_X}/E_{C_X}$ will expose the ancillary qubit to charge noise, these ancillary errors are not detrimental to the measurement due to its QND nature, which allows the measurement to be repeated multiple times.
    Note that when the $X$ measurement is not taking place and $E_J(t) \gg E_{C_\theta}$, then the $\theta$ mode is highly insensitive to the coupled ancillary qubit, meaning that errors from this qubit will not propagate to the $0$-$\pi$ qubit.

    \subsection{Capacitive disorder} \label{app:capacitive-disorder}
    In the above we assumed that all coupling capacitors are symmetric, e.g., all four capacitors coupling the $\zeta$ mode to the $Z$-measurement transmon have capacitance $C_{c_Z}$.
    Relaxing this assumption introduces spurious charge-charge couplings between the different modes of the circuit, as well as unwanted charge drives. 
    At a high level, the logical subsystem of the $0$-$\pi$ qubit will be robust to these when it is in the protected regime because these erroneous terms manifest as linear charge noise, which the $0$-$\pi$ qubit is designed to be protected against; see \cref{app:logical-subsystem}. 
    Nonetheless, below we detail the implications of these couplings for the coherence of the $0$-$\pi$ qubit and the measurement protocols.
    \begin{itemize}
        \item Internal coupling of the $0$-$\pi$ qubit modes ($\n_\theta \n_\varphi$, $\n_\theta\n_\zeta$, $\n_\varphi \n_\zeta$). 
        These are the same spurious terms that occur due to capacitive disorder within the $0$-$\pi$ qubit itself.
        Therefore, their effects are equivalent to what has already been analyzed extensively in Refs.~\cite{Dempster2014,Groszkowski2018,Paolo2019,Kolesnikow2026}:
        provided the $0$-$\pi$ qubit is in the protected regime, it is insensitive to internal capacitive disorder. 
        
        \item Spurious coupling of the $0$-$\pi$ circuit modes to the ancillary qubits ($\n_\theta \n_Z$, $\n_\varphi \n_X$, $\n_\varphi \n_Z$, $\n_\zeta \n_X$).
        The effect of the $\n_\theta \n_Z$ term will be exponentially suppressed in $\sqrt{E_J/E_{C_\theta}}$ due to the transmon-like nature of the $\theta$ mode.
        The terms coupling to the charge operators for the $\varphi$ and $\zeta$ modes could potentially lead to excitations of these modes. 
        However, in the protected regime these excitations will not lead to transitions between logical subsystems. i.e., they will be of the form
        \begin{equation}
            \ket{\mu}_L \otimes \ket{m}_S \to \ket{\mu}_L \otimes \ket{m'}_S,
        \end{equation}
        where $\ket{\mu}_L$ represents a state of the logical subsystem and $\ket{m}_S$ a state of the stabilizer subsystem.
        Excitations within the logical subsystem will be exponentially suppressed in $E_J/E_{C_\theta}$ due to the disjoint support of the logical states $\ket{\bar{0}}$ and $\ket{\bar{1}}$ in $\theta$. i.e.,
        \begin{equation}
            \basisbra{L}{0}\basisbra{S}{m} \n_\alpha \ketbasis{1}{L}\ketbasis{m'}{S} \propto e^{-\beta\sqrt{E_J/E_{C_\theta}}},
        \end{equation}
        where $\n_\alpha = \n_\theta, \n_\varphi, \n_\zeta$, and $\beta$ is a decay constant that is close to $2(2-\sqrt{2})$ in the protected regime~\cite{Kolesnikow2026}.
        See \cref{fig:subsystem-rates} for a pictorial representation of this explanation.
        
        \item Unwanted charge drives ($\eps(t) \n_\theta$, $\eps(t) \n_\varphi$, $\eps(t) \n_X$). 
        For the same reasoning as above, even though transitions in the $\theta$ and $\varphi$ modes may be driven, their effect on the logical subsystem will be exponentially suppressed in $\sqrt{E_J/E_{C_\theta}}$.
        The $\eps(t) \n_X$ term gives unwanted driving of the $X$-measurement ancillary qubit.
        This may be prevented by ensuring it is far detuned from the $\zeta$ mode frequency and the other $Z$-measurement ancillary qubit.
        
        \item Charge-charge coupling between the ancillary qubits ($\n_X \n_Z$). 
        This is standard charge-charge coupling between two unprotected qubits that may lead to a small reduction in their coherence times.
        However, the measurement protocols are already designed to work with unprotected ancillary qubits.
    \end{itemize}

    \section{Logical subsystem} \label{app:logical-subsystem}
    The $0$-$\pi$ qubit may be viewed through the lens of error correction by interpreting the logical information encoding as a subsystem rather than a subspace.
    That is, rather than simply using the two lowest-energy eigenstates to define the qubit, we decompose the full Hilbert space into a tensor product of a protected computational space that encodes the qubit, and a remaining infinite-dimensional space that we effectively trace out when making a measurement.
    This approach has been used for GKP codes, whereby this tracing operation corresponds to an ideal decoding map~\cite{Pantaleoni2020,Shaw2024}, and was also used to quantify the gate error in Ref.~\cite{Kolesnikow2026} for the $0$-$\pi$ qubit.
    In our case, the fact that this decoding is performed automatically when the qubit is measured is a consequence of the protected nature of the qubit.

    For the $0$-$\pi$ qubit, the Hilbert space for the computational modes, the $\theta$ and $\varphi$ modes, is decomposed as
    \begin{equation}
        \mathcal{H} = \mathcal{H}_L \otimes \mathcal{H}_S,
    \end{equation}
    where $\mathcal{H}_L$ is the logical subsystem, which is isomorphic to $\mathbb{C}^2$, and $\mathcal{H}_S$ is isomorphic to the infinite-dimensional Hilbert space $\mathcal{H}$.
    The latter subsystem is referred to as the \textit{stabilizer subsystem}~\cite{Shaw2024} because its basis states correspond to different stabilizer syndromes when the $0$-$\pi$ qubit is viewed as an error-correcting code.

    We can then define a set of \textit{measurement operators}, which act as the $2 \times 2$ Pauli matrices on the logical Hilbert space and the identity on the stabilizer subsystem:
    \begin{align}
        \Xm &= \hat{X} \otimes \hat{I}, \label{eqn:X-m-generic}\\
        \Ym &= \hat{Y} \otimes \hat{I}, \label{eqn:Y-m-generic}\\
        \Zm &= \hat{Z} \otimes \hat{I}. \label{eqn:Z-m-generic}
    \end{align}
    Therefore, these are the operators we wish to measure for the $0$-$\pi$ qubit, and a measurement of these corresponds to what would be obtained after an error-correction and decoding map.

    A choice of subsystem decomposition corresponds to a  particular choice of QEC decoder.
    For the $0$-$\pi$ qubit we choose to define everything closer to $\theta=0$ to be a logical $\ket{\bar0}$ state and everything closer to $\theta = \pi$ to be a logical $\ket{\bar1}$ state, regardless of its support in $\varphi$.
    This yields the following $Z$-type measurement operator
    \begin{equation} \label{eqn:Z-m}
        \Zm = \sgn[\cos\thetah],
    \end{equation}
    which is equivalent to a binning-type decoder along the $\theta$ quadrature.
    For the $X$-type operator we must account for the fact that the $\cos\thetah\cos\varphih$ potential shifts the location of the potential wells in $\varphi$ depending on the value of $\theta$.
    This can be achieved by defining the $X$-type measurement operator to be
    \begin{equation} \label{eqn:X-m}
        \Xm = \sgn [\cos(\pi \n_\theta + \pi \n_\varphi)],
    \end{equation}
    so that a shift in $\theta$ by $\pi$ is accompanied by a shift in $\varphi$ by $\pi$.
    Note that both $\Zm$ and $\Xm$ are unitary, Hermitian and anticommute.
    Finally we define $\Ym = i \Xm\Zm$ so that the three operators satisfy the Pauli algebra.
    These measurement operators define a subsystem decomposition via \cref{eqn:X-m-generic,eqn:Y-m-generic,eqn:Z-m-generic}, and decompose the basis states $\ket{\theta, n_\varphi}$ of the full $(\theta, \varphi)$ Hilbert space as follows:
    \begin{equation} \label{eqn:subsystem-mapping}
        \ket{\theta,n_\varphi} \to \left( \sgn[\cos(\pi n_\varphi)] \right)^\mu \ket{\mu} \otimes \ket*{\tilde{\theta}, n_\varphi},
    \end{equation}
    such that $\theta = \mu\pi + \tilde{\theta}$, where $\mu = 0,1$ defines the logical state and $\tilde{\theta} \in [-\pi/2,\pi/2)$ is the remainder of $\theta$ modulo $\pi$.

    \begin{figure}[b]
		\centering
        \includegraphics[scale=1]{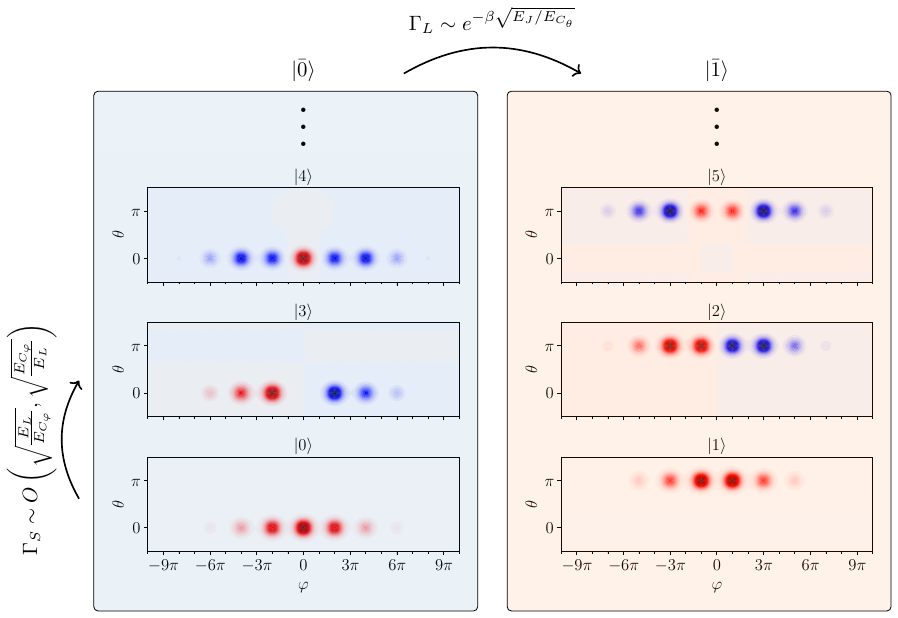}%
		\caption{\textbf{Subsystem decomposition for the $\bm{0}$-$\bm{\pi}$ qubit eigenstates.}
        Red (blue) regions denote positive (negative) values of the real part of the wavefunction.
        $\beta$ is a decay constant that is close to $2(2 - \sqrt{2})$ in the protected regime~\cite{Kolesnikow2026}.
        The $0$-$\pi$ qubit parameters are $E_{C_\theta}/E_{C_\varphi} = 0.01$, $E_J/E_{C_\varphi} = 5$ and $E_L/E_{C_\theta} = 4.1 \times 10^{-3}$.}
        \label{fig:subsystem-rates}
	\end{figure}

    \begin{figure}[t]
		\centering
        \subfloat{\label{fig:X-eigenstate-wavefns}}%
        \subfloat{\label{fig:X-meas-errors-higher-excited-states}}%
        \includegraphics[scale=1]{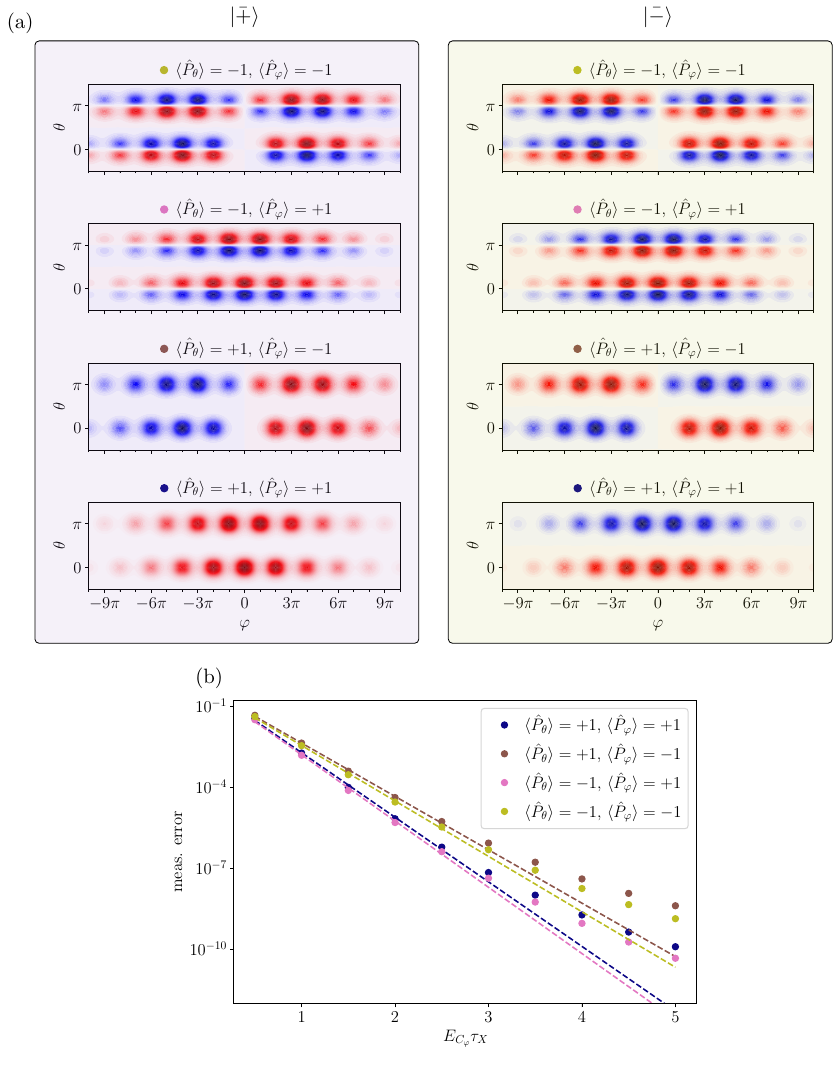}%
		\caption{\textbf{$\bm{X}$ measurement of higher excited eigenstates.} 
        \textbf{(a)} Wavefunctions for logical $\ket{\bar{\pm}}$ states in different parity sectors.
        Red (blue) regions denote positive (negative) values of the real part of the wavefunction.
        \textbf{(b)} $X$-measurement errors for the corresponding states.
        The dark blue dots represent the same data as shown for the $E_{J_\mathrm{min}} = 0$ case in \cref{fig:X-meas-errors}.
        The $0$-$\pi$ qubit parameters are $E_{C_\theta}/E_{C_\varphi} = 0.01$, $E_J/E_{C_\varphi} = 2.5$ and $E_L/E_{C_\theta} = 4.1 \times 10^{-3}$.}
        \label{fig:X-meas-different-parity-sectors}
	\end{figure}

    To show this, we compute the action of $\Zm$ and $\Xm$ on an arbitrary logical $\mu$-codeword:
    \begin{subequations}
        \begin{align}
            \ket{\psi_\mu} &= \ket{\mu} \otimes \ket*{\tilde{\psi}} \\
            &= \ket{\mu} \otimes \int_{-\pi/2}^{\pi/2} d\tilde{\theta} \int_{-\infty}^\infty d n_\varphi \, \tilde{\psi}(\tilde{\theta}, n_\varphi) \,\ket*{\tilde{\theta}, n_\varphi} \\
            &\to \int_{-\pi/2}^{\pi/2} d\tilde{\theta} \int_{-\infty}^\infty d n_\varphi \, \tilde{\psi}(\tilde{\theta}, n_\varphi)(\sgn[\cos(\pi n_\varphi)])^\mu\ket*{\mu\pi + \tilde{\theta}, n_\varphi},
        \end{align}
    \end{subequations}
    where $\tilde{\psi}(\tilde{\theta}, n_\varphi) = \braket*{\tilde{\theta}, n_\varphi}{\tilde{\psi}}$ is the stabilizer-state wavefunction.
    From the final expression, it is easy to see that $\sgn[\cos\thetah]\ket{\psi_\mu} = (-1)^\mu \ket{\mu} \otimes \ket*{\tilde{\psi}}$, and thus $\sgn[\cos\thetah]$ acts as $\hat{Z} \otimes \hat{I}$.
    To verify that $\sgn[\cos(\pi\n_\theta + \pi \n_\varphi)]$ acts as $\hat{X} \otimes \hat{I}$, note that because $\theta$ is periodic and $n_\theta$ is discrete, then $\sgn[\cos(\pi \n_\theta)] = e^{i\pi \n_\theta}$, and so
    \begin{subequations}
        \begin{align}
            \sgn[\cos(\pi \n_\theta + \pi\n_\varphi)] \ket{\psi_\mu} &= \int_{-\pi/2}^{\pi/2} d\tilde{\theta} \int_{-\infty}^\infty d n_\varphi \, \tilde{\psi}(\tilde{\theta}, n_\varphi)(\sgn[\cos(\pi n_\varphi)])^{\mu + 1 \, \mathrm{mod} \, 2}\ket*{(\mu + 1 \,\mathrm{mod}\,2)\pi + \tilde{\theta}, n_\varphi} \\
            &\to \ket{\mu + 1 \, \mathrm{mod} \, 2} \otimes \int_{-\pi/2}^{\pi/2} d\tilde{\theta} \int_{-\infty}^\infty d n_\varphi \, \tilde{\psi}(\tilde{\theta}, n_\varphi) \,\ket*{\tilde{\theta}, n_\varphi} \\
            &= \ket{\mu + 1 \, \mathrm{mod} \, 2} \otimes \ket*{\tilde{\psi}}.
        \end{align}
    \end{subequations}
    Thus, the operators specified in \cref{eqn:Z-m,eqn:X-m} act as $\hat{Z} \otimes \hat{I}$ and $\hat{X} \otimes \hat{I}$ under the decomposition specified by \cref{eqn:subsystem-mapping}.

    To see why defining the logical states in terms of the subsystem decomposition rather than using the two lowest energy eigenstates is important, consider the example eigenstates for a well-protected $0$-$\pi$ qubit plotted in \cref{fig:subsystem-rates}.
    Each eigenstate belongs to a different half of the logical subsystem depending on whether it is localized near $\theta = 0$ or $\theta = \pi$.
    Owing to the fact that $E_J \gg E_{C_\theta}$, then every eigenstate has $|\expval*{\Zm}|$ very close to 1 and transitions from one half of the logical subsystem to the other are exponentially suppressed in $\sqrt{E_J/E_{C_\theta}}$.
    In contrast, if we were to define the qubit to be the lowest-energy two-dimensional subspace, then transitions outside of this subspace are not necessarily exponentially suppressed in $\sqrt{E_J/E_{C_\theta}}$.
    For example, the $\ket{3}$ state consists of a single excitation of the $\varphi$ mode localized near $\theta = 0$, and therefore a transition from $\ket{0} \to \ket{3}$ due to an operator that is polynomial $\varphih$ or $\n_\varphi$ would scale only polynomially in $E_L/E_{C_\varphi}$. 
    We note that dephasing processes are exponentially suppressed in both the subspace and subsystem pictures since the splitting between the energies in each doublet in \cref{fig:subsystem-rates} is exponentially small in $\sqrt{E_J/E_{C_\theta}}$.

    \subsection{Measuring higher excited states} \label{app:measuring-higher-excited-states}
    
    In the above discussion we explained how states outside of the lowest-energy manifold should still be considered logical states. 
    Here we show that our measurement scheme indeed yields low measurement errors when measuring higher excited eigenstates of the $0$-$\pi$ qubit.
    Since our simulations for the $Z$-basis measurement treat $\theta$ as a binary variable, it is clear that so long as $\theta$ is localized around $0$ or $\pi$, then our measurement scheme will yield low error rates regardless of the energy of the state.
    In our simulations for the $X$-basis measurement, we used the lowest-energy manifold.
    So we now show that higher excited states still result in a low measurement error for the $X$-basis measurement scheme.
    
    The two lowest energy eigenstates of the $0$-$\pi$ qubit are $+1$ eigenstates of the parity operators $\P_\theta : \theta \to -\theta$ and $\P_\varphi : \varphi  \to -\varphi$.
    Since $\P_\theta$ and $\P_\varphi$ commute with the $0$-$\pi$ qubit Hamiltonian, the eigenstates of the Hamiltonian must be eigenstates of both of these parity operators.
    In \cref{fig:X-eigenstate-wavefns} we plot the wavefunctions for four doublets corresponding to the different parity sectors of $\P_\theta$ and $\P_\varphi$, and in \cref{fig:X-meas-errors-higher-excited-states} we show their corresponding measurement errors.
    We observe that all eigenstates yield similar measurement errors, and decrease exponentially with increasing measurement time, but that the eigenstates in the $+1$ parity sector of $\P_\varphi$ decay slightly slower.

    \section{Information extraction from GKP states in the \texorpdfstring{$\zeta$}{ζ} mode} \label{app:Z-meas-error-comparison}
\begin{figure}[t]
        \centering
        \includegraphics[width=\SingleColLineWidth]{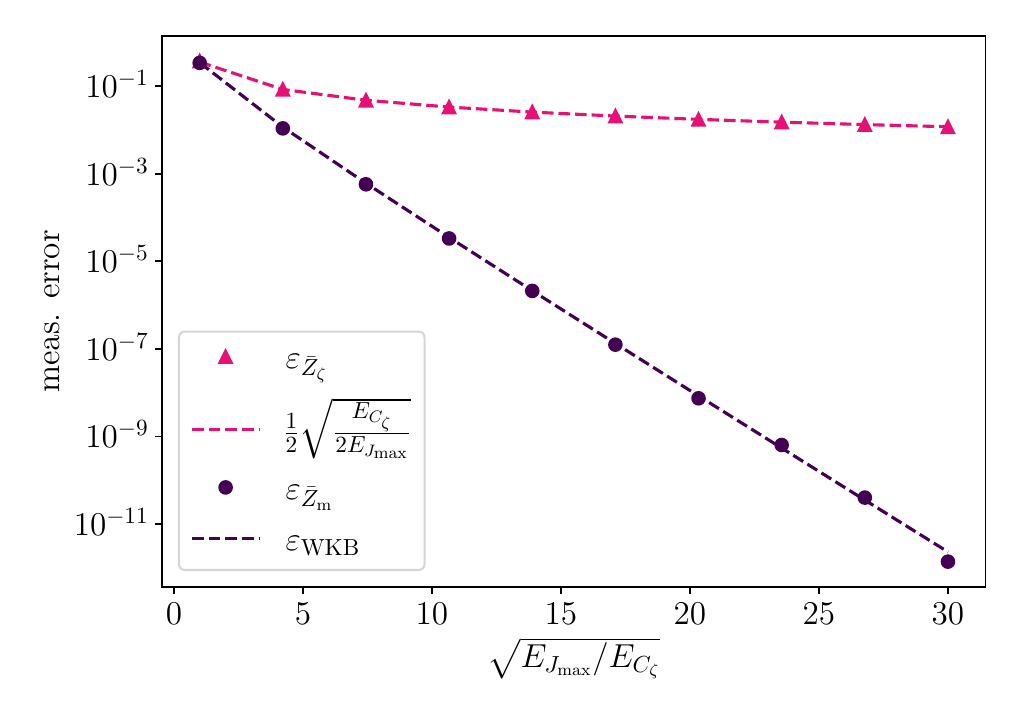}
        \caption{\textbf{Comparison of measurement errors using $\bm{\bar{Z}_\zeta}$ or $\bm{\Zm}$.}
        Using the logical GKP operator $\bar{Z}_\zeta$ to measure the qubit results in only polynomial suppression of the measurement error with respect to $E_{J_\mathrm{max}}/E_{C_\zeta}$, whereas using the measurement operator $\Zm$ results in exponential suppression of the measurement error.
        Numerical results are shown by the circle and triangle markers, whereas the analytical estimates given by \cref{eqn:meas-error-Z,eqn:meas-error-Zm-WKB} are shown in the pink and purple dashed lines, respectively.
        The $0$-$\pi$ qubit parameters are the same as in \cref{fig:Z-meas-errors}.}
        \label{fig:Z-meas-errors-comparison}
    \end{figure}

    In this appendix we discuss different strategies to extract the logical information contained in the GKP state that has been prepared in the $\zeta$ mode during the $Z$-basis measurement.
    The $Z$-basis information is stored in the position-basis wavefunction of the GKP state.
    Similarly to the case of the protected qubit, the measurement operator for the GKP code is $\Zm = \sgn [\cos\zetah]$.
    Ideally, this could be be extracted from a single homodyne measurement of the $\zeta$-mode oscillator.
    However, in practice, performing a homodyne measurement of the $\zeta$-mode oscillator would require it to be lossy, which may affect the coherence of the $0$-$\pi$ qubit and its logical operations~\cite{Paolo2019,Kolesnikow2026}.
    Instead, the information can be extracted from a dispersively coupled ancillary qubit after performing entangling operations between the oscillator and the ancillary qubit.
    Specifically, a single application of an ECD gate followed by a measurement of the ancillary qubit can be used to obtain $\expval*{e^{i \zetah}}$, which corresponds to a measurement of $\bar{Z}_\zeta$, the logical Pauli-$Z$ operator for the GKP code.
    However, this leads to an error rate that scales only polynomially in $\sqrt{E_{J_\mathrm{max}}/E_{C_\zeta}}$ (equivalently, in the squeezing of the GKP state) rather than exponentially; see \cref{fig:Z-meas-errors-comparison}.
    Instead, multiple ECD gates may be performed to retrieve the expectation value of $\Zm$.
    This may be seen by writing out the Fourier series for $\Zm$:
    \begin{equation}
        \Zm = \sgn[\cos\zetah] = \frac{1}{\pi} \sum_{n \in \Z} \frac{(-1)^n}{n+1/2} e^{i(2n+1)\zetah}.
    \end{equation}
    Since the expectation value of each $e^{i(2n+1)\zetah}$ term may be obtained from an ECD gate followed by a measurement of the ancillary qubit, multiple measurements of the oscillator state can be made, and the expectation value of $\Zm$ can be reconstructed through post-processing~\cite{Matsos2024,Matsos2025}.
    Alternatively, it was shown recently that $\Zm$ can be measured in a single shot by using Quantum Signal Processing techniques~\cite{Chalermpusitarak2026}, whereby a series of ECD gates are performed followed by a single measurement on the ancillary qubit.
    There also exist other readout strategies for measuring finite-energy GKP states in oscillator-qubit systems using multiple ECD gates followed by a single measurement of the qubit~\cite{Singh2025}.
    However, the errors for these also scale only polynomially in the squeezing of the GKP state and would therefore lead to the measurement error of the $0$-$\pi$ qubit scaling polynomially in $\sqrt{E_{J_\mathrm{max}}/E_{C_\zeta}}$. 
    
    We now compute measurement errors for $\expval*{\bar{Z}_\zeta}$ and $\expval*{\Zm}$ analytically, showing that the former scales as $\sqrt{E_{C_\zeta}/2E_{J_\mathrm{max}}}$, whereas the latter scales exponentially in the inverse of this ratio.
    In our calculations we set $\hbar = 1$ for notational convenience.
    The prepared GKP states in the $\zeta$ mode consist of a superposition of cosine-well ground states localized at the even or odd multiples of $\pi$ and are weighted by an envelope function that is Gaussian with a variance given by the harmonic oscillator impedance $Z_\zeta$.
    To simplify the calculations, we ignore the envelope because it has no contribution to the error rate to leading order in $1/Z_\zeta$.
    Using the quadratic approximation to the cosine well, the ground-state wavefunctions at each cosine-potential minima are a Gaussian with variance $\sqrt{8E_{C_\zeta}/E_{J_\mathrm{max}}}$:
    \begin{equation} \label{eqn:Gaussian-wavefn}
        \psi(\zeta) = \left(\frac{E_{J_\mathrm{max}}}{8 \pi^2 E_{C_\zeta}}\right)^{1/8} \exp\left( -\frac{1}{2} \sqrt{\frac{E_{J_\mathrm{max}}}{8 E_{C_\zeta}}} \zeta^2 \right).
    \end{equation}
    The expectation value of the Gaussian localized at the $n$-th cosine well with $e^{i \zetah}$ is $(-1)^n e^{-\sqrt{E_{C_\zeta} / 2E_{J_\mathrm{max}}}}$.
    Thus, the measurement error with respect to $\bar{Z}_\zeta$ is
    \begin{subequations}
        \begin{align}
            \eps_{\bar{Z}_\zeta} &= \frac{1}{2} - \frac{\expval*{\bar{Z}_\zeta}_0 - \expval*{\bar{Z}_\zeta}_1}{4} \\
            &\approx \frac{1}{2} - \frac{1}{2}e^{-\sqrt{E_{C_\zeta} / 2E_{J_\mathrm{max}}}} \\
            &\approx \frac{1}{2}\sqrt{\frac{E_{C_\zeta}}{2 E_{J_\mathrm{max}}}}, \label{eqn:meas-error-Z}
        \end{align}
    \end{subequations}
    where the approximation holds when $E_{J_\mathrm{max}} \gg E_{C_\zeta}$, and the subscript $0$ ($1$) represents the expectation value with the logical $\ket{\bar 0}$ ($\ket{\bar 1}$) state. 
    Therefore, the measurement error associated with measuring $\bar{Z}_\zeta$ is only polynomially suppressed in $E_{J_\mathrm{max}}/E_{C_\zeta}$; see the pink triangles (numerically evaluated measurement error) and pink dashed line [\cref{eqn:meas-error-Z}] in \cref{fig:Z-meas-errors-comparison}.

    To estimate the error associated with measuring $\Zm$ we note that $\expval*{\sgn[\cos\zetah]}$ corresponds to subtracting the probability of the state being closer to the odd multiples of $\pi$ from the probability of it being closer to the even multiples of $\pi$.
    Therefore, the error rate for measuring a logical $0$ or $1$ state is given by the probability that the wavefunction occupies the regions closer to even or odd multiples of $\pi$:
    \begin{equation}
        \eps_\mu = \sum_{n \in \Z} \int_{(2n+\mu)\pi + \pi/2}^{(2n+\mu)\pi + 3\pi/2}  d\zeta \, |\psi_\mu(\zeta)|^2.
    \end{equation}
    Since we neglect the envelope of the GKP states, their wavefunctions are perfectly periodic and the logical error rate for both states can be estimated by simply calculating the amount of the cosine-well ground state that occupies the region between $\zeta = \pi/2$ and $\zeta = \pi$:  
    \begin{equation} \label{eqn:p-mu-wavefn-estimate}
        \eps = 2\int_{\pi/2}^\pi d\zeta \, |\psi(\zeta)|^2.
    \end{equation}
    Because this error estimate depends on the ground-state wavefunction far from the center of the cosine well, the Gaussian approximation to the wavefunction is insufficient to accurately capture the error.
    Instead, we estimate the wavefunction in the classically forbidden region, $\zeta >\pi/2$, using a WKB calculation~\cite{Dunne2017,Benderskii1995}.
    We start by assuming that sufficiently far away from $\zeta = 0$ the wavefunction decays exponentially: 
    \begin{equation}
        \psi(\zeta) = e^{-S(\zeta)/\lambda},
    \end{equation}
    where $S(\zeta)$ is a position-dependent action that we expand perturbatively in the small ratio $\lambda = \sqrt{8 E_{C_\zeta}/E_{J_\mathrm{max}}}$:
    \begin{equation}
        S(\zeta) = S_0(\zeta) + \lambda S_1(\zeta) + \dots .
    \end{equation}
    To find the zeroth- and first-order terms we substitute the ansatz wavefunction into the Schr\"odinger equation:
    \begin{equation}
        -\frac{\lambda^2}{2}\frac{d^2 \psi(\zeta)}{d\zeta^2} + 2\sin^2(\zeta/2) \psi(\zeta) = \frac{E_0}{E_{J_\mathrm{max}}} \psi(\zeta),
    \end{equation}
    where $E_0$ is the ground-state energy, and we have taken the potential energy to be $E_{J_\mathrm{max}}(1 - \cos\zeta) = 2E_{J_\mathrm{max}} \sin^2(\zeta/2)$.
    This gives 
    \begin{align}
        \left( \frac{dS_0}{d\zeta} \right)^2 = 4 \sin^2\left(\frac{\zeta}{2}\right), \\
        \frac{d S_1}{d\zeta} \frac{dS_0}{d\zeta} = \frac{1}{2} \frac{d^2S_0}{d\zeta^2} - \nu,
    \end{align}
    where $\nu = E_0/\sqrt{8 E_{C_\zeta} E_{J_\mathrm{max}}}$.
    The solution to these coupled differential equations is
    \begin{align}
        S_0(\zeta) &= 4[1 - \cos(\zeta/2)], \\
        S_1(\zeta) &= \frac{1}{2}\log|2\sin(\zeta/2)| - \nu \log|\tan(\zeta/4)| - \log N,
    \end{align}
    where $N$ is an integration constant.
    Thus, the ansatz wavefunction becomes
    \begin{equation} \label{eqn:WKB-wavefn}
        \psi(\zeta) = \frac{N}{2\cos(\zeta/4)} \tan^{\nu - 1/2}(\zeta/4) e^{-4[1 - \cos(\zeta/2)]/\lambda}.
    \end{equation}
    To find the normalization constant $N$, we match the ansatz solution to the Gaussian solution in the region $\sqrt{\lambda} \ll \zeta \ll \pi/2$ by Taylor expanding \cref{eqn:WKB-wavefn} and matching to \cref{eqn:Gaussian-wavefn}.
    This yields $\nu = 1/2$ and $N = 2 /(\pi\lambda)^{1/4}$, and thus the WKB wavefunction is
    \begin{equation}
        \psi(\zeta) = \left(\frac{1}{\pi\lambda} \right)^{1/4} \frac{e^{-4[1 - \cos(\zeta/2)]/\lambda}}{\cos(\zeta/4)}.
    \end{equation}
    Estimating the wavefunction at $\zeta > \pi/2$ by $\psi(\pi/2 + \delta) \approx \psi(\pi/2) e^{-\sqrt{2} \delta/\lambda}$, and extending the upper bound in \cref{eqn:p-mu-wavefn-estimate} to $\infty$, yields the error probability
    \begin{subequations}
        \begin{align}
            \eps_\mathrm{WKB} &= 2|\psi(\pi/2)|^2\int_{0}^\infty  d \delta \, e^{-2\sqrt{2}\delta/\lambda} \\
            &= 2(2 - \sqrt{2}) \sqrt{\frac{\lambda}{2\pi}} e^{-4(2 - \sqrt{2})/\lambda} \\
            &= \frac{4(\sqrt{2} - 1)}{\sqrt{\pi}} \left(\frac{E_{C_\zeta}}{2E_{J_\mathrm{max}}} \right)^{1/4} e^{-(2 - \sqrt{2})\sqrt{2E_{J_\mathrm{max}}/E_{C_\zeta}}}. \label{eqn:meas-error-Zm-WKB}
        \end{align}
    \end{subequations}
    We find good agreement between this analytical estimate (purple dashed line in \cref{fig:Z-meas-errors-comparison}) and the numerical results (purple circles in \cref{fig:Z-meas-errors-comparison}).

    \section{State preparation}
    Here we explain how to use the measurement protocols described in the main text to fault tolerantly prepare logical states the $Z$ and $X$ bases.
    
    \subsection{State preparation in the logical \texorpdfstring{$Z$}{Z} basis} \label{app:Z-state-prep}
    Once a GKP state has been prepared in the $\zeta$ mode by turning on the interaction term $-E_{J_\mathrm{int}}(t)\cos(\thetah + \zetah)$, its logical information is transferred to the ancillary qubit.
    If this information is entirely transferred to the ancillary qubit and a perfect measurement of the qubit is made, then a measurement outcome of $0$ or $1$ projects the $0$-$\pi$ qubit into a logical $\ket{\bar{0}}$ or $\ket{\bar{1}}$ state. 
    To compensate for measurement errors due to ancillary qubit noise, the $\zeta$ mode can be cooled and the measurement procedure can be repeated multiple times.
    The majority vote on the outcomes then signals which state has been prepared.
    
    As stressed in the main text and \cref{app:Z-meas-error-comparison}, the extent to which the logical information from the $\zeta$ mode is mapped onto the ancillary qubit depends on the exact ECD gate sequence that is implemented. 
    Different ECD gate sequences will lead to a different state preparation fidelity.
    To see this, assume that the $0$-$\pi$ qubit begins in an arbitrary logical state $\alpha \ket{\bar 0} + \beta \ket{\bar 1}$.
    After the interaction term has been turned on, the state of the combined $0$-$\pi$ qubit and $\zeta$ mode system is
    \begin{equation}
        \ket{\Psi} = \alpha \ket{\bar 0}_{0\mathrm{-}\pi} \ket{\bar0}_\zeta + \beta \ket{\bar1}_{0\mathrm{-}\pi} \ket{\bar1}_\zeta,
    \end{equation}
    where $\ket{\bar\mu}_{0\mathrm{-}\pi}$ denotes a logical $\mu \in \{0,1\}$ state encoded across the $\theta$ and $\varphi$ modes of the $0$-$\pi$ circuit, and $\ket{\bar\mu}_\zeta$ denotes the finite-energy GKP codeword prepared in the $\zeta$ mode of the $0$-$\pi$ circuit.
    A particular ECD gate sequence leads to the following measurement operators for an outcome of $0$ or $1$ on the ancillary qubit
    \begin{align}
        M_0 &= \sqrt{1 - \eta} \ketbasisbra{\bar 0}{\zeta}{\bar 0} + \sqrt{\eta'} \ketbasisbra{\bar 1}{\zeta}{\bar 1},\\
        M_1 &= \sqrt{\eta} \ketbasisbra{\bar 0}{\zeta}{\bar 0} + \sqrt{1-\eta'} \ketbasisbra{\bar 1}{\zeta}{\bar 1},
    \end{align}
    where $\eta = 1 - \basisbra{\zeta}{\bar 0}\bar{Z}_\mathrm{ECD}\ketbasis{\bar 0}{\zeta}$ and $\eta' = 1 + \basisbra{\zeta}{\bar 1}\bar{Z}_\mathrm{ECD}\ketbasis{\bar 1}{\zeta}$ are the expectation values of the finite-energy GKP states with the measurement operator that is implemented via the ECD gate sequence (e.g., $\bar{Z}_\mathrm{ECD} = e^{i\zetah}$ and $\bar{Z}_\mathrm{ECD} = \sgn[\cos\zetah]$ were the two examples we studied in \cref{app:Z-meas-error-comparison}).
    The preparation infidelity due to this imperfect measurement is then
    \begin{align}
        F_0 &= \basisbra{0-\pi}{\bar 0} \Tr_\zeta\left[ \frac{M_0 \ketbra{\Psi} M_0^\dag}{\bra{\Psi} M_0^\dag M_0 \ket{\Psi}} \right] \!\! \ketbasis{\bar 0}{{0-\pi}} = \frac{\abs*{\alpha}^2(1 - \eta)}{\abs*{\alpha}^2(1 - \eta) + \abs*{\beta}^2 \eta'}, \\
        F_1 &= \basisbra{0-\pi}{\bar 1} \Tr_\zeta\left[ \frac{M_1 \ketbra{\Psi} M_1^\dag}{\bra{\Psi} M_1^\dag M_1 \ket{\Psi}} \right] \!\! \ketbasis{\bar 1}{{0-\pi}} = \frac{\abs*{\beta}^2(1 - \eta')}{\abs*{\alpha}^2 \eta + \abs*{\beta}^2(1 - \eta')}.
    \end{align}
    For non-zero $\alpha, \beta$ and $1 - \eta$, $1 - \eta'$, the infidelities to first order in $\eta,\eta'$ are
    \begin{align}
        1 - F_0 &\approx \abs{\frac{\beta}{\alpha}}^2 \eta', \\
        1 - F_1 &\approx \abs{\frac{\alpha}{\beta}}^2 \eta.
    \end{align}
    Therefore, the preparation fidelity scales in the same way as the measurement error for the same measurement operator. 
    For example, the preparation infidelity will be exponentially suppressed in $\sqrt{E_{J_\mathrm{max}}/E_{C_\zeta}}$ if $\sgn[\cos\zetah]$ is implemented.
    
    \subsection{State preparation in the logical \texorpdfstring{$X$}{X} basis} \label{app:X-state-prep}
    After turning off the internal Josephson junctions of the $0$-$\pi$ qubit and measuring the charge of the $\theta$ mode, a logical $\ket{\bar{\pm}}$ state of the qubit can be prepared by turning the Josephson junction back on via the inverse ramp.
    A $\ket{\bar{+}}$ ($\ket{\bar{-}}$) state will be prepared if the measured charge had even (odd) parity.
    Since, in general, the states after the turn-off have support on multiple different charge eigenstates with the same parity, the state that is prepared in the $0$-$\pi$ qubit after turning $E_J$ back on will depend on the exact charge state that was measured.
    Whilst they will all be $\pm 1$ eigenstates of $\Xm$, they may have support on higher excited eigenstates of the $0$-$\pi$ qubit Hamiltonian.
    Due to the subsystem encoding, this is not fundamentally an issue for operating the $0$-$\pi$ qubit in the fully protected regime.
    Nonetheless, to avoid preparing higher excited states, either the measurement can be repeated until a charge of $n_\theta = 0$ or $n_\theta = 1$ is measured, or the turn-off can be performed more slowly to ensure that the system remains in the lowest-energy doublet throughout the evolution; see \cref{fig:charge-state-adiabaticity} and the discussion in \cref{app:adiabatic-turn-off} for details on the adiabaticity constraints for remaining in the lowest-energy doublet.

    \section{Numerical simulations} \label{app:numerical-simulations}
    Here we explain our numerical approach to simulate the measurement protocols.
    We make use of the split-operator method detailed in Appendix B of Ref.~\cite{Kolesnikow2026} for time-dynamics state-vector simulations.
    In this section we set $\hbar = 1$ for convenience.
    
    The split-operator method exploits the fact that we can partition the Hamiltonian into position and momentum components, $\H = \H_x + \H_p$, and Trotterize the unitary time-evolution operator via
    \begin{equation}
        \U(dt) = \U_x(dt/2) \U_p(dt) \U_x(dt/2) + O(dt^3),
    \end{equation}
    where $\U(t) = e^{-i(\H_x + \H_p) t}$, $\U_{x/p}(dt) = e^{-i \H_{x/p} t}$.
    Time-evolution is then obtained by repeated multiplication of the position or momentum unitary with the state vector in the position or momentum basis, interspersed with Fourier transforms:
    \begin{equation} \label{eqn:split-op-step}
        \psi(t + dt) = \U_x(dt/2) \mathcal{F}^{-1} \U_p(dt) \mathcal{F} \U_x(dt/2) \psi(t),
    \end{equation}
    where $\psi(t)$ denotes the position wavefunction at time $t$ and $\mathcal{F}$ and $\mathcal{F}^{-1}$ denote the Fourier and inverse Fourier transforms.

    \subsection{\texorpdfstring{$Z$}{Z} measurement}
    For the simulations of the $Z$ measurement in \cref{fig:Z-meas-errors}, we simulate the Hamiltonian
    \begin{equation} \label{eqn:H-Z}
        \H(t) = 4 E_{C_\zeta} \n_\zeta^2 + E_L \zetah^2 \pm E_{J_\mathrm{int}}(t) \cos\zetah,
    \end{equation}
    which describes the $\zeta$ mode oscillator acted on by the interaction term $\pm E_{J_\mathrm{int}}(t)\cos(\zetah + \thetah)$, where we use $-$ for $\ket{\bar{0}}$ and $+$ for $\ket{\bar{1}}$.
    This model treats the qubit $\bar{Z}_\mathrm{m}$ eigenstate as a classical number, which is justified in the limit that $E_J \gg E_{C_\theta}$ and the qubit states are well localized near $\theta = 0$ and $\theta = \pi$.
    The interaction amplitude is given by the error-function pulse:
    \begin{equation}
        E_{J_\mathrm{int}}(t) = \frac{E_{J_\mathrm{max}}}{2} \left[ \erf\left( \frac{t - t_0^\uparrow}{\tau_Z^\uparrow} \right) - \erf \left( \frac{t - t_0^\downarrow}{\tau_Z^\downarrow} \right) \right],
    \end{equation}
    where $\tau_Z^\uparrow = 1/\omega_\zeta$ and $\tau_Z^\downarrow$ are the pulse turn-on and turn-off times, and $t_0^\uparrow = 2\pi \tau_Z^\uparrow$ and $t_0^\downarrow = 3 t_0^\uparrow$ are the centers of each ramp.
    
    The initial state is the ground state of the harmonic oscillator
    \begin{equation} \label{eqn:Gaussian-initial-state}
        \psi(0) = \mathcal{N}_\sigma(0),
    \end{equation}
    where $\mathcal{N}_\sigma(x)$ denotes a Gaussian with mean $x$ and standard deviation $\sigma$.
    Here $\sigma^2 = \sqrt{4E_{C_\zeta}/E_L}$ is the variance of the harmonic oscillator ground state.
    We then evolve this state under \cref{eqn:H-Z} with $-$ ($+$) using the split-operator method, and compute the expectation value of the final state with $\bar{Z}_\mathrm{ECD}$ to obtain $\expval*{\bar{Z}_\mathrm{ECD}}_0$ ($\expval*{\bar{Z}_\mathrm{ECD}}_1$), where $\bar{Z}_\mathrm{ECD}$ is the operator on the $\zeta$ mode that is implemented through the ECD gate sequence; see \cref{app:Z-meas-error-comparison}. 
    The measurement error is then computed as 
    \begin{equation}
        \varepsilon_Z = \frac{1}{2} - \frac{\expval*{\bar{Z}_\mathrm{ECD}}_0 - \expval*{\bar{Z}_\mathrm{ECD}}_1}{4}.
    \end{equation}

    To test the validity of treating the qubit state as a classical number, as in equation \cref{eqn:H-Z}, we also simulate $Z$ measurements using an effective model in which the protected qubit is dynamical, but has a reduced state-space allowing for practical simulations. 
    In these simulations we use the Hamiltonian~\cite{Kolesnikow2026}
    \begin{equation} \label{eqn:H-alpha-full}
        \H(t) = 4 E_{C_\zeta} \n_\zeta^2 + E_L \zetah^2 + 4 E_{C_\alpha} \n_\alpha^2 - E_{J_\alpha} \cos\alphah - E_{J_{2\alpha}} \cos 2\alphah - E_{J_\mathrm{int}}(t) \cos(\alphah - \pi \n_\alpha), 
    \end{equation}
    where $\alpha$ is the effective degree of freedom.
    An explanation of the effective model and the result of $Z$ measurement simulations using it are discussed in \cref{app:Z-meas-eff-model}.
    Since the last term in \cref{eqn:H-alpha-full} involves an interaction between the position of one mode and the momentum of another we modify the split-operator step to perform the Fourier transforms in each mode separately~\cite{Kolesnikow2026}.
    The initial states in this case are different, and are given by
    \begin{align} 
        \psi_0(0) &= \mathcal{N}_\sigma(0) \otimes \frac{1}{\sqrt{N}}, \label{eqn:seed-state-0}\\
        \psi_1(0) &= \mathcal{N}_\sigma(0) \otimes \sqrt{\frac{2}{N}}\cos(\alpha), \label{eqn:seed-state-1}
    \end{align}
    where $\mathcal{N}_\sigma(0)$ is the same Gaussian on the $\zeta$ mode as in \cref{eqn:Gaussian-initial-state} and $N$ are the number of basis points for the $\alpha$ mode used in the simulation.
    The states $1/\sqrt{N}$ and $\sqrt{2/N}\cos(\alpha)$ are states with support on only even or odd values of $n_\alpha$, which correspond to $\pm 1$ eigenstates of $\Zm$ as per the effective model; see \cref{app:Z-meas-eff-model} and Ref.~\cite{Kolesnikow2026}. 

    \subsection{\texorpdfstring{$X$}{X} measurement}
    For the $X$ measurement, we simulate the Hamiltonian
    \begin{equation} \label{eqn:H-X}
        \H(t) = 4 E_{C_\theta} \n_\theta^2 + 4 E_{C_\varphi} \n_\varphi^2 + E_L \varphih^2 - 2E_J(t) \cos\thetah\cos\varphih.
    \end{equation}
    Here $E_J(t)$ is given by the error-function ramp
    \begin{equation}
        E_J(t) = E_J - \frac{E_J - E_{J_\mathrm{min}}/2}{2} \left[ 1 + \erf\left( \frac{t - t_0}{\tau_X^\downarrow} \right) \right],
    \end{equation}
    where $\tau_X^\downarrow$ is the turn-off time and $t_0 = 2\pi \tau_X^\downarrow$ is the center of the ramp.
    To find the eigenstates of $\H(0)$ we use imaginary time evolution by performing the split-operator method with $dt \to -i dt$ .
    We use a seed state that has support on even or odd multiples of $n_\varphi$ and then evolve for long enough so that the state relaxes into the symmetric or antisymmetric superposition of the lowest energy states localized near $\theta= 0$ and $\theta = \pi$.
    Note that whilst the true ground state may be some other superposition of these two localized states, we exploit the fact that tunneling between these two regions occurs on a much longer timescale because $E_{C_\theta} \ll E_{C_\varphi}$.
    Thus, we can reliably prepare a desired superposition, determined by the choice of seed state, by choosing the total evolution time to be on the order of $1/E_{C_\varphi}\ll 1/E_{C_\theta}$. 
    
    We choose the seed states to be the same as in \cref{eqn:seed-state-0,eqn:seed-state-1} where now the first mode corresponds to the $\varphi$ mode and the second mode corresponds to the $\theta$ mode.
    Note that both of these are even-valued functions of both $\theta$ and $\varphi$.
    In \cref{app:measuring-higher-excited-states} we considered higher excited states that are odd-valued functions of $\theta$ and/or $\varphi$.
    For an odd-valued function of $\varphi$ we multiply the Gaussian seed state by $\varphi$ and normalize, and for an odd-valued function of $\theta$ we replace $\cos\alpha$ in \cref{eqn:seed-state-1} with $\sin2\varphi$ or $\sin\varphi$ for a $+1$ or $-1$ eigenstate of $\Xm$, respectively. 
    To prepare these higher excited states we must also perform a projection onto the corresponding parity sector before each split-operator step in \cref{eqn:split-op-step} via 
    \begin{equation}
        \psi(\varphi, \theta) \to \mathcal{P}_\pm^\varphi \mathcal{P}_\pm^\theta \psi(\varphi,\theta),
    \end{equation}
    where $\psi(\varphi,\theta)$ is the position wavefunction and 
    \begin{align}
        \mathcal{P}_\pm^\varphi \psi(\varphi,\theta) &= \frac{1}{2}\left[\psi(\varphi,\theta) \pm \psi(-\varphi,\theta) \right], \\
        \mathcal{P}_\pm^\theta \psi(\varphi,\theta) &= \frac{1}{2}\left[\psi(\varphi,\theta) \pm \psi(\varphi,-\theta) \right].
    \end{align}
    Once we have prepared an eigenstate of $\H(0)$, we evolve it under \cref{eqn:H-Z} and take the expectation value of the final state with $e^{-i\pi \n_\theta}$.
    Doing this for both of the evolved states, we compute the measurement error as
    \begin{equation}
        \varepsilon_X = \frac{1}{2} - \frac{\expval*{e^{-i \pi \n_\theta}}_+ - \expval*{e^{-i \pi \n_\theta}}_-}{4},
    \end{equation}
    where the subscript $\pm$ denotes an expectation value taken with respect to the time-evolved $\pm 1$ eigenstate of $\Xm$.

    \section{\texorpdfstring{$Z$}{Z}-measurement simulation with effective model} \label{app:Z-meas-eff-model}
    For the numerical simulations of the $Z$-basis measurement in the main text we treated $\theta$ as a classical binary variable that takes values $\theta = 0$ or $\theta = \pi$.
    This is justified in the protected regime where $E_J \gg E_{C_\theta}$, since the $0$-$\pi$ qubit states will be well-localized near $\theta=0$ and $\theta = \pi$.
    Furthermore, since $\Zm = \sgn[\cos\thetah]$, the quantity of interest for the $Z$-basis measurement is independent of the $\varphi$ mode.
    Here we will discuss an alternative model and compare the results.
    
    To account for both the $\theta$ and $\varphi$ modes of the $0$-$\pi$ qubit whilst maintaining a small enough state-space for the numerical simulations, we may reduce the two-mode Hamiltonian in \cref{eqn:H-theta-phi} to a single-mode effective Hamiltonian, using the model presented in Ref.~\cite{Kolesnikow2026}:
    \begin{equation} \label{eqn:H-alpha}
        \H_\alpha = 4 E_{C_\alpha} \n_\alpha^2 - E_{J_\alpha} \cos\alphah - E_{J_{2\alpha}} \cos 2\alphah,
    \end{equation}
    where $[\alphah, \n_\alpha] = i$. 
    Here $\alpha$ represents an effective degree of freedom capturing the dynamics of both the $\theta$ and $\varphi$ mode.
    Its conjugate momentum $n_\alpha$ can be thought of as indexing the minima of the two-dimensional cosine $-2E_J\cos\theta\cos\varphi$ located at $(\theta, \varphi) = (\mu\pi, m\varphi)$, where $\mu = 0,1$ and $m$ is an integer such that $\mu + m$ is even.         
    Note that \cref{eqn:H-alpha} corresponds to the Hamiltonian for an ideal single-mode protected qubit [\cref{eqn:H-ideal-qubit}] that is perturbed by a $2\pi$-periodic term $\cos\alphah$.
    This has the effect of further splitting the the two ground state energies.
    The effective energies $E_{C_\alpha}, E_{J_\alpha}, E_{J_{2\alpha}}$ depend on the energies for the $0$-$\pi$ qubit $E_{C_\theta}, E_{C_\varphi}, E_J, E_L$.
    In particular, the ratio $E_{J_\alpha}/E_{J_{2\alpha}}$ dictates how protected the $0$-$\pi$ qubit is, and is exponentially suppressed in the charging energy ratio $\sqrt{E_{C_\varphi}/E_{C_\theta}}$;
    for $E_J/E_{C_\varphi} = 5$ and charging energy ratios $E_{C_\varphi}/E_{C_\theta}$ between 50 and 200, $E_{J_\alpha}/E_{J_{2\alpha}}$ ranges between $10^{-4}$ and $10^{-11}$~\cite{Kolesnikow2026}.

\begin{figure}[h]
        \centering
        \includegraphics[width=\SingleColLineWidth]{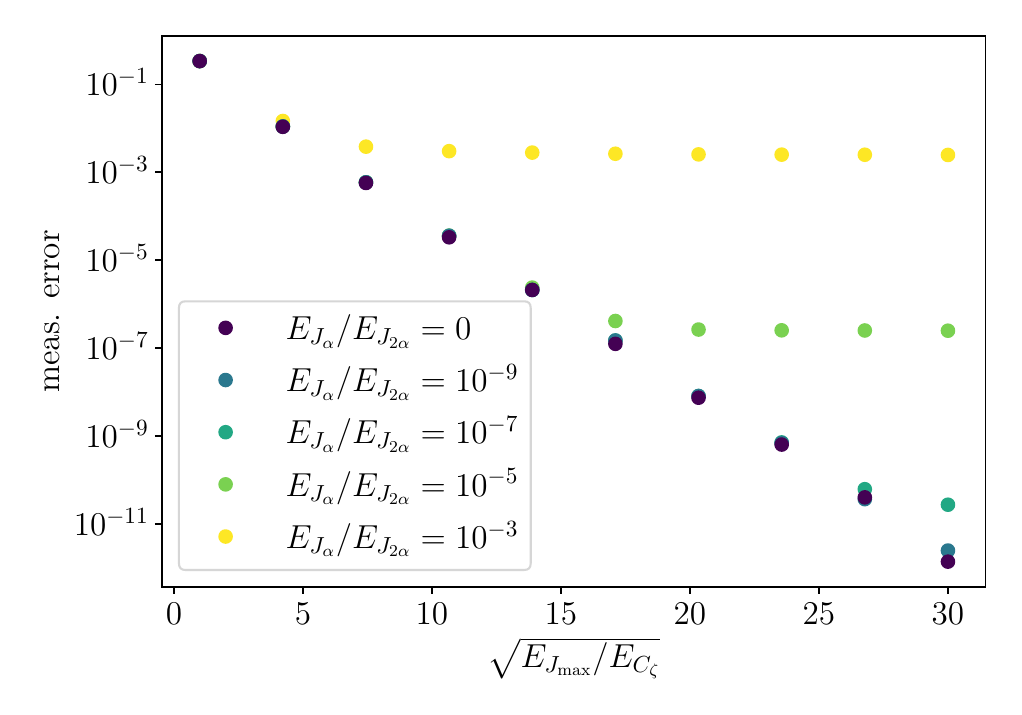}
        \caption{\textbf{$\bm{Z}$ measurement errors with effective model.}
        The $E_{J_\alpha}/E_{J_{2\alpha}} = 0$ data is identical to the $\tau_Z^\downarrow = 0$ data in \cref{fig:Z-meas-errors}, where $\theta$ is treated as a classical bit.
        Throughout, $E_{J_{2\alpha}}/E_{C_\alpha} = 100$, $E_{C_\alpha} = E_{C_\zeta}$, $E_L/E_{C_\zeta} = 4.1 \times 10^{-3}$ and $\tau_Z^\downarrow = 0$.}
        \label{fig:Z-meas-errors-eff-model}
    \end{figure}

    In \cref{fig:Z-meas-errors-eff-model} we compare the measurement errors using the effective model to the ones presented in \cref{fig:Z-meas-errors} where $\theta$ was treated classically.
    We find that for $E_{J_\alpha}/E_{J_{2\alpha}} = 10^{-9}$ there is negligible difference to the case where $\theta$ is treated classically, but that a significant error plateau occurs at values of $E_{J_\alpha}/E_{J_{2\alpha}} \ge 10^{-5}$.
    Given $E_{J_\alpha}/E_{J_{2\alpha}} \ge 10^{-5}$ in the regions where the $0$-$\pi$ qubit is not well-protected anyway, the measurement will likely not be the rate-limiting error process.

    \section{Adiabatic \texorpdfstring{$E_J$}{EJ} turn-off} \label{app:adiabatic-turn-off}

    Here we give additional details on the adiabaticity of the turn-off for the internal Josephson elements required for the $X$-basis measurement.
    Owing to the subsystem encoding, a key distinction between the measurement error and pure adiabaticity is that the measurement error is only concerned with transitions that lead to population in different parity sectors of $\expval{e^{i \pi \n_\theta}}$ for the final states, whereas pure adiabaticity forbids any transition.
    Whilst the $\ket{n_\theta = 0}$ and $(\ket{n_\theta=+1} + \ket{n_\theta=-1})/\sqrt{2}$ eigenstates of the final Hamiltonian are adiabatically linked to the lowest energy logical $\ket{\bar{+}}$ and logical $\ket{\bar{-}}$ states of the $0$-$\pi$ qubit, transitions to higher excited states with the same value of $\expval*{\Xm}$ during the turn-off do not affect the measurement error.
    For example, the final wavefunctions depicted in \cref{fig:X-meas-protocol} are not the $\ket{n_\theta = 0}$ and $(\ket{n_\theta=+1} + \ket{n_\theta=-1})/\sqrt{2}$ states, yet they incur a vanishing measurement error.

    To investigate this further we estimate the adiabatic condition and compare its scaling to the measurement error.
    From the adiabatic theorem, a rough condition for adiabaticity is that~\cite{Zwiebach2022}
    \begin{equation} \label{eqn:adiabaticity-1}
        \abs{\frac{\bra{m} \partial H / \partial t \ket{n}}{(E_m - E_n)^2}} \ll 1
    \end{equation}
    must hold throughout the adiabatic process, where $\ket{m}$, $\ket{n}$ are the instantaneous eigenstates and $E_m$, $E_n$ are their eigenenergies.
    The Hamiltonian we are concerned with is \cref{eqn:H-theta-phi} with $E_J \to E_J(t)$.
    At the beginning of the turn-off when $E_J(t)$ is large, the eigenstates of this Hamiltonian are nearly eigenstates of the turn-off term $-2E_J(t)\cos\thetah \cos\varphih$ and thus transitions will be strongly suppressed.
    At the end of the turn-off when $E_J(t) \ll E_{C_\theta}$, the instantaneous eigenstates are well approximated by the product states $\ketbasis{n}{\theta} \otimes\ketbasis{m}{\varphi}$, where $\ketbasis{n}{\theta}$ are the $\theta$-mode charge states with energy $4 E_{C_\theta} n_\theta^2$ and $\ketbasis{m}{\varphi}$ are $\varphi$-mode harmonic oscillator eigenstates with energy $4m_\varphi\sqrt{E_{C_\varphi} E_L}$.
    In the regime that $E_L/E_{C_\varphi} > (E_{C_\theta}/E_{C_\varphi})^2$, then the $\varphi$-mode transitions will be larger than the $\theta$-mode transitions and so we can can treat the $\varphi$ mode to be in its ground state~\footnote{Note that this is equivalent to making a Born-Oppenheimer approximation. This approximation is not usually valid for the $0$-$\pi$ qubit since $E_J \gtrsim E_{C_\varphi}$ in the protected regime~\cite{Kolesnikow2026}, but here we are interested in the unprotected case when $E_J \ll E_{C_\varphi}$.}.
    This results in the effective Hamiltonian for the $\theta$-mode dynamics
    \begin{equation} \label{eqn:H-eff}
        \H_\mathrm{eff} = 4E_{C_\theta} \n_\theta^2 + \H^{(1)} + \H^{(2)} + \dots,
    \end{equation}
    where $\H^{(1)}$ and $\H^{(2)}$ represent the first- and second-order corrections.
    Note that this effective model is only valid because $E_J(t) \ll E_{C_\varphi}$ and differs to the one used in \cref{app:Z-meas-eff-model} where $E_J \gtrsim E_{C_\varphi}$.
    The corrections in \cref{eqn:H-eff} are found using perturbation theory and may be simplified as follows:
    \begin{subequations}
        \begin{align}
            \H_0^{(1)} &= \basisbra{\varphi}{0} \hat{V} \ketbasis{0}{\varphi} \\
            &= -2E_J(t)\basisbra{\varphi}{0}\cos\varphih\ketbasis{0}{\varphi}\cos\thetah \\
            &= -2E_J(t) e^{-\pi Z_\varphi / 2R_Q}  \cos\thetah, \label{eqn:E01-2}\\
            \H_0^{(2)} &= \sum_{m=1}^\infty \frac{\abs*{\basisbra{\varphi}{m} \hat{V} \ketbasis{0}{\varphi}}^2}{m\omega_\varphi} \\
            &=-\frac{4E_J(t)^2}{\hbar \omega_\varphi}\sum_{m=1}^\infty \frac{\abs*{\basisbra{\varphi}{m} \cos\varphih \ketbasis{0}{\varphi}}^2}{m} \cos^2\thetah\\
            &= -\frac{4E_J(t)^2}{\hbar \omega_\varphi} e^{-\pi Z_\varphi/R_Q} \sum_{k=1}^\infty \frac{(\pi Z_\varphi / R_Q)^{2k}}{2k(2k)!} \cos^2\thetah \label{eqn:E02-2}\\
            &\sim -\frac{2E_J(t)^2 R_Q}{\pi \hbar \omega_\varphi Z_\varphi} \cos^2\thetah \label{eqn:E02-3}\\
            &= -\frac{E_J(t)^2}{2 E_{C_\varphi}} \cos^2\thetah, \label{eqn:E01-4}
        \end{align}
    \end{subequations}
    where $\ket{m}$ and $Z_\varphi/R_Q = \sqrt{ E_{C_\varphi}/\pi^2 E_L}$, with $R_Q = h/4e^2$ the superconducting resistance quantum, are the eigenstates and impedance for the unperturbed harmonic oscillator, respectively.
    In \cref{eqn:E02-2} we used the following matrix element expressions
    \begin{subequations} \label{eqn:cos-phi-mat-elts}
    \begin{align} 
        \basisbra{\varphi}{2k}\cos\varphih\ketbasis{0}{\varphi} &= \frac{-e^{-\pi Z_\varphi/2R_Q} (\pi Z_\varphi/R_Q)^{k}}{\sqrt{(2k)!}}, \\
        \basisbra{\varphi}{2k+1}\cos\varphih\ketbasis{0}{\varphi} &= 0,
    \end{align}
    \end{subequations}
    and in \cref{eqn:E02-3} we used the following asymptotic equivalence
    \begin{equation}
        e^{-x} \sum_{k=1}^\infty \frac{x^{2k}}{(2k)!} \sim \frac{1}{2x}, \qquad x \to \infty,
    \end{equation}
    which is justified because $Z_\varphi/R_Q \gg 1$. 
    Therefore, the effective Hamiltonian in \cref{eqn:H-eff} to second-order in $E_J$ is
    \begin{equation} \label{eqn:H-eff-1}
        \H_\mathrm{eff} = 4 E_{C_\theta} \n_\theta^2 - 2E_J(t) e^{-\pi Z_\varphi / 2R_Q}  \cos\thetah - \frac{E_J(t)^2}{4 E_{C_\varphi}} \cos2\thetah,
    \end{equation}
    where we used that $2\cos^2\thetah = \cos2\thetah + 1$ and dropped constant terms.
    Plugging the effective Hamiltonian [\cref{eqn:H-eff-1}] into the adiabatic condition [\cref{eqn:adiabaticity-1}] and using that $\basisbra{\theta}{n}\cos\thetah\ketbasis{n \pm 1}{\theta} = \basisbra{\theta}{n}\cos2\thetah\ketbasis{n \pm 2}{\theta} = 1/2$ yields
    \begin{equation} \label{eqn:adiabaticity-2}
        \frac{E_J'(t_*)}{16 E_{C_\theta}^2}e^{-\pi Z_\varphi/2R_Q} +\frac{[E_J'(t_*)]^2}{2048 E_{C_\varphi} E_{C_\theta}^2} \ll 1,
    \end{equation}
    where $t_*$ is the time at which $E_J(t_*) \ll E_{C_\theta}$.
    The first term is the parity changing term as it leads to $\ketbasis{n}{\theta} \to \ketbasis{n \pm 1}{\theta}$, but it is exponentially suppressed in the impedance $Z_\varphi$.
    Thus, at the end of the turn-off, the dominant transitions will be the parity conserving ones $\ketbasis{n}{\theta} \to \ketbasis{n \pm 2}{\theta}$ induced by the last term in \cref{eqn:H-eff-1}, which is not exponentially suppressed in $Z_\varphi$.
    
\begin{figure}
    \centering
        \includegraphics[width=1.7\SingleColLineWidth]{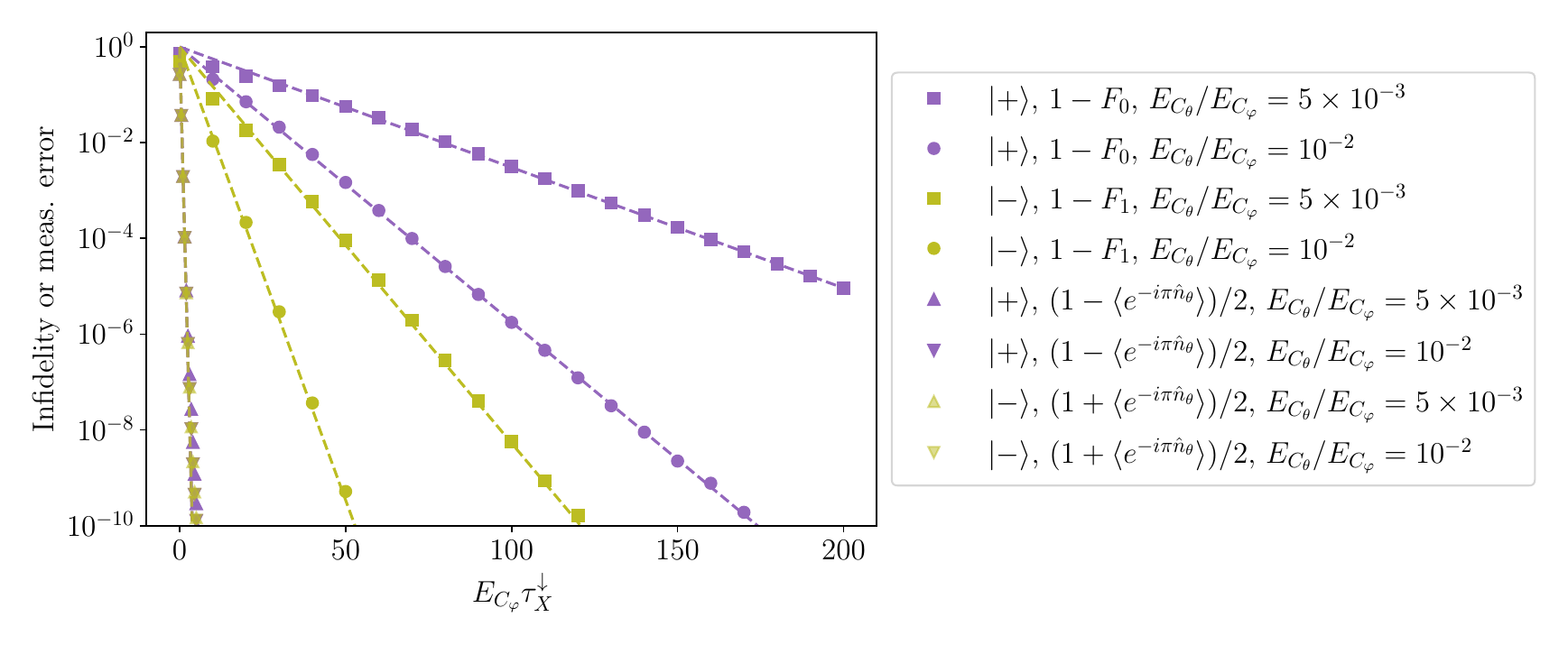}
        \caption{\textbf{Comparison of $\bm{X}$ measurement errors to charge state infidelities.} 
        Purple (green) represents the errors for the initial $\ket{\bar{+}}$ ($\ket{\bar{-}}$) state. 
        The purple (green) squares and circles represents the infidelity of the final state after the $E_J$ turn-off with the $\ketbasis{0}{\theta}$ [$(\ketbasis{+1}{\theta} + \ketbasis{-1}{\theta})/\sqrt{2}$] state.
        The purple (green) triangles represent the measurement error $1 - \expval{e^{i \pi \n_\theta}}$ ($1 + \expval{e^{i \pi \n_\theta}}$).
        Averaging the values for the purple and green downwards triangles ($E_{C_\theta}/E_{C_\varphi} = 0.01$) recovers the blue data in \cref{fig:X-meas-errors}.
        Throughout, $E_L/E_{C_\theta} = 4.1 \times 10^{-3}$, $E_{J}(0)/E_{C_\varphi} = 2.5$ and $E_{J_\mathrm{min}} = 0$.}
        \label{fig:charge-state-adiabaticity}
    \end{figure}

    In \cref{fig:charge-state-adiabaticity} we compare the measurement error for the $\ket{\bar{\pm}}$ states with the infidelities of the final states with the $\ketbasis{0}{\theta}$ or $(\ketbasis{+1}{\theta} + \ketbasis{-1}{\theta})/\sqrt{2}$ states at two different charging energy ratios $E_{C_\theta}/E_{C_\varphi}$.
    We see that the measurement errors decrease much more rapidly with measurement time than than the infidelities, consistent with the fact that the transitions affecting the measurement error are heavily suppressed compared to the transitions that affect the fidelity in the regime of large $Z_\varphi/R_Q$. 
    \Cref{fig:charge-state-adiabaticity} also shows how the infidelities improve significantly when $E_{C_\theta}/E_{C_\varphi}$ is doubled, whereas the measurement errors remain nearly invariant.
    The improvement in the infidelities with $E_{C_\theta}/E_{C_\varphi}$ is consistent with the $E_{C_\theta}$ denominators in \cref{eqn:adiabaticity-2}.
    The fact that the measurement errors are nearly invariant with $E_{C_\theta}/E_{C_\varphi}$ indicates that the rate-limiting transitions for the measurement occur in the middle of the turn-off where $E_J/E_{C_\theta}$ is small enough so that the instantaneous eigenstates are not close to the eigenstates of $\cos\thetah \cos\varphih$, but large enough so that they are not well approximated by the product states $\ketbasis{n}{\theta} \otimes \ketbasis{m}{\varphi}$. 
    
    \section{Effect of impedance on measurement errors} \label{app:impedance}
    Here we investigate how the impedance of the $0$-$\pi$ qubit affects the measurement errors.
    The $0$-$\pi$ qubit has a single inductive energy that should be small in order for the qubit to be protected.
    In contrast, there are multiple different charging energies in the $0$-$\pi$ qubit, and therefore there are different impedances for each of the modes.
    For the measurement errors, the two relevant impedances are $Z_\zeta$ and $Z_\varphi$, defined by $Z_\zeta \pi/ R_Q =\sqrt{E_{C_\zeta}/E_L}$ and $Z_\varphi \pi/ R_Q = \sqrt{E_{C_\varphi}/E_L}$, where $R_Q = h/4e^2$ is the superconducting resistance quantum.
    These two impedances are related by the charging energy ratio via $Z_\varphi/Z_\zeta = \sqrt{E_{C_\varphi}/E_{C_\zeta}}$.
    In the protected regime where $E_{C_\theta} \ll E_{C_\varphi}$, we have that $E_{C_\zeta} \approx E_{C_\theta}$.
    In Ref.~\cite{Kolesnikow2026} it was shown that a charging energy ratio of $E_{C_\theta}/E_{C_\varphi} = 10^{-2}$ and impedance of $Z_\zeta/R_Q = 5$ (equivalently $Z_\varphi/R_Q \approx 50$) are sufficient for the $0$-$\pi$ qubit to have a fully protected single and two-qubit phase gate.
    We use these values for our simulations to estimate the measurement errors in the main text.

\begin{figure}[t]
        \centering
        \includegraphics[width=1.7\SingleColLineWidth]{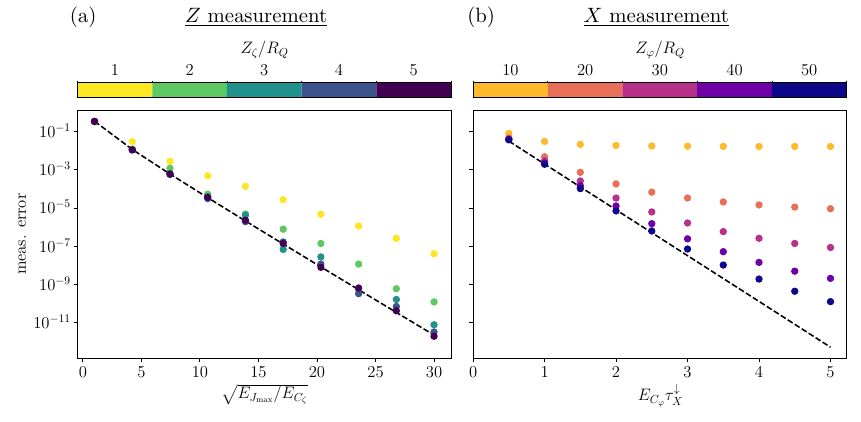}
        \caption{\textbf{Impact of impedance on measurement errors.}
        \textbf{(a)} $Z$ measurement errors as a function of the maximum Josephson energy at different impedances of the $\zeta$ mode.
        The $Z_\zeta/R_Q = 5$ (equivalentally, $E_L/E_{C_\zeta} = 4.1 \times 10^{-3}$) data corresponds to the $\tau_Z^\downarrow = 0$ data in \cref{fig:Z-meas-errors}.
        All other parameters are the same as in \cref{fig:Z-meas-errors}.
        \textbf{(b)} $X$ measurement errors as a function of the turn-off time at different impedances of the $\varphi$ mode.
        The $Z_\varphi/R_Q = 50$ (equivalentally, $E_L/E_{C_\theta} = 4.1 \times 10^{-3}$) data corresponds to the $E_{J_\mathrm{min}} = 0$ data in \cref{fig:X-meas-errors}.
        All other parameters are the same as in \cref{fig:X-meas-errors}.
        }
        \label{fig:Impedance}
    \end{figure}

    Here we examine how the measurement errors change when we reduce the impedance.
    \Cref{fig:Impedance} reveals that, in general, decreasing the impedance leads to larger error rates.
    However, in the case of the $Z$-basis measurement, the measurement errors are largely unchanged over the range of $\sqrt{E_{J_\mathrm{max}}/E_{C_\zeta}}$ plotted here until $Z_\zeta/R_Q\le 2$.
    This is because the expectation value of $\Zm$ is independent of the envelope of the prepared GKP states to leading order; see \cref{app:Z-meas-error-comparison}.
    In contrast, the $X$-basis measurement errors exhibit a clear trend with $Z_\varphi/R_Q$, whereby the divergence from the exponential decay occurs sooner at smaller $Z_\varphi/R_Q$.
    This is due to the fact that parity-changing transitions are exponentially suppressed in $Z_\varphi$; see \cref{app:adiabatic-turn-off}.  

	\bibliography{refs}

@inproceedings{Aharonov1997,
    author = {Aharonov, D. and Ben-Or, M.},
    title = {{Fault-tolerant quantum computation with constant error}},
    year = {1997},
    isbn = {0897918886},
    publisher = {Association for Computing Machinery},
    address = {New York, NY, USA},
    url = {https://doi.org/10.1145/258533.258579},
    doi = {10.1145/258533.258579},
    booktitle = {Proceedings of the Twenty-Ninth Annual ACM Symposium on Theory of Computing},
    pages = {176},
    numpages = {13},
    location = {El Paso, Texas, USA},
    series = {STOC '97}
}

@misc{Aramburu2026,
    author = {Pablo Aramburu Sanchez and Trevyn F. Q. Larson and Anthony P. McFadden and Constantin Schrade and Joshua Combes and Andr{\'a}s Gyenis},
    title = {{Revisiting the multi-mode rhombus circuit as a biased-noise qubit}},
    year = {2026},
    eprint = {2605.06430},
    archivePrefix={arXiv},
    primaryClass={quant-ph},
    url={https://arxiv.org/abs/2605.06430}, 
}

@InProceedings{Beauseigneur2026,
  author    = {Brieuc Beauseigneur and Louis Lattier and Aron Vanselow and Zaki Leghtas and Philippe Campagne-Ibarcq},
  booktitle = {APS Global Physics Summit Abstracts},
  title     = {{Towards Dissipative Stabilization of Four-Legged Cat Qubits in Superconducting Circuits: Part II}},
  year      = {2026},
  month     = mar,
  pages     = {MAR.W14.3},
  volume    = {2026},
  eid       = {MAR.W14.3},
  url       = {https://summit.aps.org/smt/2026/events/MAR-W14/3},
    address = {Denver, CO, USA}
}

@article{Benderskii1995,
    title = {{Tunneling splittings in model 2D potentials. III. Generalization to N-dimensional case}},
    journal = {Chem. Phys.},
    volume = {198},
    number = {3},
    pages = {281},
    year = {1995},
    issn = {0301-0104},
    doi = {https://doi.org/10.1016/0301-0104(95)00138-E},
    url = {https://www.sciencedirect.com/science/article/pii/030101049500138E},
    author = {V.A. Benderskii and S.Yu. Grebenshchikov and G.V. Mil'nikov},
}

@article{Brooks2013,
	title = {{Protected gates for superconducting qubits}},
	author = {Brooks, Peter and Kitaev, Alexei and Preskill, John},
	journal = {Phys. Rev. A},
	volume = {87},
	issue = {5},
	pages = {052306},
	numpages = {26},
	year = {2013},
	month = {May},
	publisher = {American Physical Society},
	doi = {10.1103/PhysRevA.87.052306},
	url = {https://link.aps.org/doi/10.1103/PhysRevA.87.052306}
}

@Article{CampagneIbarcq2020,
  author    = {P. Campagne-Ibarcq and A. Eickbusch and S. Touzard and E. Zalys-Geller and N. E. Frattini and V. V. Sivak and P. Reinhold and S. Puri and S. Shankar and R. J. Schoelkopf and L. Frunzio and M. Mirrahimi and M. H. Devoret},
  journal   = {Nature},
  title     = {{Quantum error correction of a qubit encoded in grid states of an oscillator}},
  year      = {2020},
  month     = aug,
  number    = {7821},
  pages     = {368--372},
  volume    = {584},
  doi       = {10.1038/s41586-020-2603-3},
  publisher = {Springer Science and Business Media {LLC}},
}

@misc{Chalermpusitarak2026,
    author = {Teerawat Chalermpusitarak and Ting Rei Tan},
    title = {{Measuring Bosonic Observables via Bosonic Quantum Signal Processing}},
    note = {in preparation},
}

@article{Ciaccia2024,
  author    = {Ciaccia, Carlo and Haller, Roy and Drachmann, Asbj{\o}rn C. C. and Lindemann, Tyler and Manfra, Michael J. and Schrade, Constantin and Sch{\"o}nenberger, Christian},
  journal   = {Commun. Phys.},
  title     = {{Charge-4e supercurrent in a two-dimensional InAs-Al superconductor-semiconductor heterostructure}},
  year      = {2024},
  issn      = {2399-3650},
  month     = Jan,
  number    = {1},
  pages = {41},
  volume    = {7},
  doi       = {10.1038/s42005-024-01531-x},
  publisher = {Springer Science and Business Media LLC},
}

@article{Dempster2014,
  title = {{Understanding degenerate ground states of a protected quantum circuit in the presence of disorder}},
  author = {Dempster, Joshua M. and Fu, Bo and Ferguson, David G. and Schuster, D. I. and Koch, Jens},
  journal = {Phys. Rev. B},
  volume = {90},
  issue = {9},
  pages = {094518},
  numpages = {12},
  year = {2014},
  month = {Sep},
  publisher = {American Physical Society},
  doi = {10.1103/PhysRevB.90.094518},
  url = {https://link.aps.org/doi/10.1103/PhysRevB.90.094518}
}

@InProceedings{Dunne2017,
    author={Dunne, Gerald V.
    and {\"U}nsal, Mithat},
    editor={Fauvet, Fr{\'e}d{\'e}ric
    and Manchon, Dominique
    and Marmi, Stefano
    and Sauzin, David},
    title={{WKB and resurgence in the Mathieu equation}},
    booktitle={Resurgence, Physics and Numbers},
    year={2017},
    publisher={Scuola Normale Superiore},
    address={Pisa},
    pages={249--298},
    isbn={978-88-7642-613-1}
}

@Article{Fluhmann2018,
  author    = {C. Fl{\"u}hmann and V. Negnevitsky and M. Marinelli and J.P. Home},
  journal   = {Phys. Rev. X},
  title     = {{Sequential Modular Position and Momentum Measurements of a Trapped Ion Mechanical Oscillator}},
  year      = {2018},
  month     = apr,
  number    = {2},
  pages     = {021001},
  volume    = {8},
  doi       = {10.1103/physrevx.8.021001},
  publisher = {American Physical Society ({APS})},
}

@article{Gottesman2001,
	title = {{Encoding a qubit in an oscillator}},
	author = {Gottesman, Daniel and Kitaev, Alexei and Preskill, John},
	journal = {Phys. Rev. A},
	volume = {64},
	issue = {1},
	pages = {012310},
	numpages = {21},
	year = {2001},
	month = {Jun},
	publisher = {American Physical Society},
	doi = {10.1103/PhysRevA.64.012310},
	url = {https://link.aps.org/doi/10.1103/PhysRevA.64.012310}
}

@article{Groszkowski2018,
    doi = {10.1088/1367-2630/aab7cd},
    url = {https://doi.org/10.1088/1367-2630/aab7cd},
    year = {2018},
    month = {apr},
    publisher = {IOP Publishing},
    volume = {20},
    number = {4},
    pages = {043053},
    author = {Groszkowski, Peter and Paolo, A Di and Grimsmo, A L and Blais, A and Schuster, D I and Houck, A A and Koch, Jens},
    title = {{Coherence properties of the $0$-$\pi$ qubit}},
    journal = {New J. Phys.},
}

@Article{Guo2024,
  author    = {Guo, Guo-Liang and Leng, Han-Bing and Liu, Xin},
  journal   = {New J. Phys.},
  title     = {{Parity-spin superconducting qubit based on topological insulators}},
  year      = {2024},
  issn      = {1367-2630},
  month     = jun,
  number    = {6},
  pages     = {063005},
  volume    = {26},
  doi       = {10.1088/1367-2630/ad4b58},
  publisher = {IOP Publishing},
}

@misc{Guo2026,
    author = {Guo-Liang Guo},
    title = {{Full Gate-Voltage Control of a Parity-Protected Superconducting Qubit with an Altermagnetic Josephson Junction}},
    year = {2026},
    eprint = {2607.04097},
    archivePrefix={arXiv},
    primaryClass={quant-ph},
    url={https://arxiv.org/abs/2607.04097}, 
}

@article{Gyenis2021,
  title = {{Experimental Realization of a Protected Superconducting Circuit Derived from the $0$-$\pi$ Qubit}},
  author = {Gyenis, Andr{\'a}s and Mundada, Pranav S. and Di Paolo, Agustin and Hazard, Thomas M. and You, Xinyuan and Schuster, David I. and Koch, Jens and Blais, Alexandre and Houck, Andrew A.},
  journal = {PRX Quantum},
  volume = {2},
  issue = {1},
  pages = {010339},
  numpages = {18},
  year = {2021},
  month = {Mar},
  publisher = {American Physical Society},
  doi = {10.1103/PRXQuantum.2.010339},
  url = {https://link.aps.org/doi/10.1103/PRXQuantum.2.010339}
}

@article{Gyenis2021a,
  title = {{Moving beyond the Transmon: Noise-Protected Superconducting Quantum Circuits}},
  author = {Gyenis, Andr{\'a}s and Di Paolo, Agustin and Koch, Jens and Blais, Alexandre and Houck, Andrew A. and Schuster, David I.},
  journal = {PRX Quantum},
  volume = {2},
  issue = {3},
  pages = {030101},
  numpages = {15},
  year = {2021},
  month = {Sep},
  publisher = {American Physical Society},
  doi = {10.1103/PRXQuantum.2.030101},
  url = {https://link.aps.org/doi/10.1103/PRXQuantum.2.030101}
}

@phdthesis{Hassani2024,
    title={{Superconducting qubits capable of dynamic switching between protected and high-speed control regimes}},
    author={Hassani, Farid},
    year={2024},
    school={Institute of Science and Technology Austria},
    url={https://doi.org/10.15479/at:ista:17133}
}

@misc{Hays2025,
  author        = {Hays, Max and Kim, Junghyun and Oliver, William D.},
  title         = {{Non-degenerate noise-resilient superconducting qubit}},
  year          = {2025},
  month         = feb,
  archiveprefix = {arXiv},
  copyright     = {arXiv.org perpetual, non-exclusive license},
  doi           = {10.48550/ARXIV.2502.15459},
  eprint        = {2502.15459},
  primaryclass  = {quant-ph},
  publisher     = {arXiv},
}

@article{Kalashnikov2020,
  title = {{Bifluxon: Fluxon-Parity-Protected Superconducting Qubit}},
  author = {Kalashnikov, Konstantin and Hsieh, Wen Ting and Zhang, Wenyuan and Lu, Wen-Sen and Kamenov, Plamen and Di Paolo, Agustin and Blais, Alexandre and Gershenson, Michael E. and Bell, Matthew},
  journal = {PRX Quantum},
  volume = {1},
  issue = {1},
  pages = {010307},
  numpages = {15},
  year = {2020},
  month = {Sep},
  publisher = {American Physical Society},
  doi = {10.1103/PRXQuantum.1.010307},
  url = {https://link.aps.org/doi/10.1103/PRXQuantum.1.010307}
}

@InProceedings{Kim2024,
  author    = {Kim, Junghyun and Rosen, Ilan and An, Junyoung and Hays, Max and Di Paolo, Agustin and Ding, Leon and Azar, Kate and Gertler, Jeffrey and Hazard, Thomas and Gingras, Michael and Niedzielski, Bethany and Stickler, Hannah and Sliwa, Katrina and Schwartz, Mollie and Yoder, Jonilyn and Orlando, Terry and Grover, Jeffrey and Serniak, Kyle and Oliver, William},
  booktitle = {APS March Meeting Abstracts},
  title     = {{Measurement of soft zero-pi qubit with parallel-plate capacitors}},
  year      = {2024},
  month     = mar,
  pages     = {M48.005},
  volume    = {2024},
  eid       = {M48.005},
  url       = {https://ui.adsabs.harvard.edu/abs/2024APS..MARM48005K},
    address = {Minneapolis, MN, USA}
}

@article{Kitaev1997,
  author    = {Kitaev, A Yu},
  journal   = {Russ. Math. Surv.},
  title     = {{Quantum computations: algorithms and error correction}},
  year      = {1997},
  issn      = {1468-4829},
  month     = dec,
  number    = {6},
  pages     = {1191},
  volume    = {52},
  doi       = {10.1070/rm1997v052n06abeh002155},
  publisher = {Steklov Mathematical Institute},
}

@misc{Kitaev2006,
    title={{Protected qubit based on a superconducting current mirror}}, 
    author={Alexei Kitaev},
    year={2006},
    eprint={cond-mat/0609441},
    archivePrefix={arXiv},
    primaryClass={cond-mat.mes-hall}
}

@misc{Kitaev2006a,
  author = {Kitaev, Alexei},
  title  = {{Protected Qubits using Josephson Junctions}},
  year   = {2006},
  note   = {{Talk presented at the KITP Conference on Topological Phases and Quantum Computation, Kavli Institute for Theoretical Physics, University of California, Santa Barbara, 15--19 May 2006}},
  url    = {https://online.kitp.ucsb.edu/online/qubit_c06/kitaev/}
}

@article{Knill1998,
    title={{Resilient quantum computation: error models and thresholds}},
    volume={454},
    ISSN={1471-2946},
    url={http://dx.doi.org/10.1098/rspa.1998.0166},
    number={1969},
    journal={Proc. R. Soc. A},
    publisher={The Royal Society},
    author={Knill, Emanuel and Laflamme, Raymond and Zurek, Wojciech H.},
    year={1998},
    month=jan, 
    pages={365} 
}

@article{Kolesnikow2026,
  title = {{Protected Phase Gate for the 0-$\ensuremath{\pi}$ Qubit using its Internal Modes}},
  author = {Kolesnikow, Xanda C. and Smith, Thomas B. and Thomsen, Felix and Alase, Abhijeet and Doherty, Andrew C.},
  journal = {PRX Quantum},
  volume = {7},
  issue = {1},
  pages = {010306},
  numpages = {40},
  year = {2026},
  month = {Jan},
  publisher = {American Physical Society},
  doi = {10.1103/bywc-cx98},
  url = {https://link.aps.org/doi/10.1103/bywc-cx98}
}

@misc{Kumar2024,
      title={{Protomon: A Multimode Qubit in the Fluxonium Molecule}}, 
      author={Shashwat Kumar and Xinyuan You and Xanthe Croot and Tianpu Zhao and Danyang Chen and Sara Sussman and Anjali Premkumar and Jacob Bryon and Jens Koch and Andrew A. Houck},
      year={2024},
      eprint={2411.16648},
      archivePrefix={arXiv},
      primaryClass={quant-ph},
      url={https://arxiv.org/abs/2411.16648}, 
}

@article{Larsen2020,
  title = {{Parity-Protected Superconductor-Semiconductor Qubit}},
  author = {Larsen, T. W. and Gershenson, M. E. and Casparis, L. and Kringh{\o}j, A. and Pearson, N. J. and McNeil, R. P. G. and Kuemmeth, F. and Krogstrup, P. and Petersson, K. D. and Marcus, C. M.},
  journal = {Phys. Rev. Lett.},
  volume = {125},
  issue = {5},
  pages = {056801},
  numpages = {6},
  year = {2020},
  month = {Jul},
  publisher = {American Physical Society},
  doi = {10.1103/PhysRevLett.125.056801},
  url = {https://link.aps.org/doi/10.1103/PhysRevLett.125.056801}
}

@article{Le2019,
  title = {{Doubly nonlinear superconducting qubit}},
  author = {Le, Dat Thanh and Grimsmo, Arne and M{\"u}ller, Clemens and Stace, T. M.},
  journal = {Phys. Rev. A},
  volume = {100},
  issue = {6},
  pages = {062321},
  numpages = {15},
  year = {2019},
  month = {Dec},
  publisher = {American Physical Society},
  doi = {10.1103/PhysRevA.100.062321},
  url = {https://link.aps.org/doi/10.1103/PhysRevA.100.062321}
}

@article{Leblanc2025,
  author    = {Leblanc, Axel and Tangchingchai, Chotivut and Sadre Momtaz, Zahra and Kiyooka, Elyjah and Hartmann, Jean-Michel and Gustavo, Fr{\'e}d{\'e}ric and Thomassin, Jean-Luc and Brun, Boris and Schmitt, Vivien and Zihlmann, Simon and Maurand, Romain and Dumur, {\'E}tienne and De Franceschi, Silvano and Lefloch, Fran{\c}ois},
  journal   = {Nat. Commun.},
  title     = {{Gate- and flux-tunable $\sin(2\varphi)$ Josephson element with planar-Ge junctions}},
  year      = {2025},
  issn      = {2041-1723},
  month     = Jan,
  pages = {1010},
  number    = {1},
  volume    = {16},
  doi       = {10.1038/s41467-025-56245-7},
  publisher = {Springer Science and Business Media LLC},
}

@misc{Leroux2023,
    title = {{Cat-qubit-inspired gate on $\cos(2\theta)$ qubits}},
    author = {Catherine Leroux and Alexandre Blais},
    year = {2023},
    eprint = {2304.02155},
    archivePrefix={arXiv},
    primaryClass={quant-ph},
    url={https://arxiv.org/abs/2304.02155},
}

@article{Lieu2025,
  title = {{Viewing protected superconducting qubits through the lens of the cat qubit}},
  author = {Lieu, Simon and Rosenfeld, Emma L. and Noh, Kyungjoo and Hann, Connor T.},
  journal = {Phys. Rev. A},
  volume = {112},
  issue = {2},
  pages = {022414},
  numpages = {12},
  year = {2025},
  month = {Aug},
  publisher = {American Physical Society},
  doi = {10.1103/p8fy-qdc8},
  url = {https://link.aps.org/doi/10.1103/p8fy-qdc8}
}

@article{Matsos2024,
  title = {{Robust and Deterministic Preparation of Bosonic Logical States in a Trapped Ion}},
  author = {Matsos, V. G. and Valahu, C. H. and Navickas, T. and Rao, A. D. and Millican, M. J. and Kolesnikow, X. C. and Biercuk, M. J. and Tan, T. R.},
  journal = {Phys. Rev. Lett.},
  volume = {133},
  issue = {5},
  pages = {050602},
  numpages = {7},
  year = {2024},
  month = {Jul},
  publisher = {American Physical Society},
  doi = {10.1103/PhysRevLett.133.050602},
  url = {https://link.aps.org/doi/10.1103/PhysRevLett.133.050602}
}

@Article{Matsos2025,
  author    = {Matsos, V. G. and Valahu, C. H. and Millican, M. J. and Navickas, T. and Kolesnikow, X. C. and Biercuk, M. J. and Tan, T. R.},
  journal   = {Nat. Phys.},
  title     = {{Universal quantum gate set for Gottesman-Kitaev-Preskill logical qubits}},
  year      = {2025},
  issn      = {1745-2481},
  month     = aug,
  number    = {10},
  pages     = {1664--1669},
  volume    = {21},
  doi       = {10.1038/s41567-025-03002-8},
  publisher = {Springer Science and Business Media LLC},
}

@misc{Messelot2026,
  author        = {S. Messelot and A. Leblanc and J.-S. Tettekpoe and F. Lefloch and Q. Ficheux and J. Renard and {\'E}. Dumur},
  title         = {{Coherence Limits in Interference-Based $\cos(2\varphi)$ Qubits}},
  year          = {2026},
  archiveprefix = {arXiv},
  eprint        = {2601.10209},
  url          = {https://arxiv.org/abs/2601.10209},
  primaryclass  = {quant-ph},
  publisher     = {arXiv},
}

@misc{Nguyen2025,
  author        = {Nguyen, Long B. and Kim, Hyunseong and Le, Dat T. and Ersevim, Thomas and Chitta, Sai P. and Chistolini, Trevor and J{\"u}nger, Christian and Smith, W. Clarke and Stace, T. M. and Koch, Jens and Santiago, David I. and Siddiqi, Irfan},
  title         = {{The superconducting grid-states qubit}},
  year          = {2025},
  month         = sep,
  archiveprefix = {arXiv},
  copyright     = {arXiv.org perpetual, non-exclusive license},
  doi           = {10.48550/ARXIV.2509.14656},
  eprint        = {2509.14656},
  primaryclass  = {quant-ph},
  publisher     = {arXiv},
}

@misc{Nguyen2025a,
  author        = {Nguyen, Minh T. P. and Shaw, Mackenzie H.},
  title         = {{Fault-Tolerant Non-Clifford GKP Gates using Polynomial Phase Gates and On-Demand Noise Biasing}},
  year          = {2025},
  month         = nov,
  archiveprefix = {arXiv},
  copyright     = {arXiv.org perpetual, non-exclusive license},
  doi           = {10.48550/ARXIV.2511.20355},
  eprint        = {2511.20355},
  primaryclass  = {quant-ph},
  publisher     = {arXiv},
}

@misc{OBrien2025,
  author        = {O'Brien, Liam and Refael, Gil and Nathan, Frederik},
  title         = {{Exponentially robust non-Clifford gate in a driven-dissipative circuit}},
  year          = {2025},
  month         = jul,
  archiveprefix = {arXiv},
  copyright     = {Creative Commons Attribution Non Commercial Share Alike 4.0 International},
  doi           = {10.48550/ARXIV.2507.19713},
  eprint        = {2507.19713},
  primaryclass  = {quant-ph},
  publisher     = {arXiv},
}

@article{Pantaleoni2020,
  title = {{Modular Bosonic Subsystem Codes}},
  author = {Pantaleoni, Giacomo and Baragiola, Ben Q. and Menicucci, Nicolas C.},
  journal = {Phys. Rev. Lett.},
  volume = {125},
  issue = {4},
  pages = {040501},
  numpages = {6},
  year = {2020},
  month = {Jul},
  publisher = {American Physical Society},
  doi = {10.1103/PhysRevLett.125.040501},
  url = {https://link.aps.org/doi/10.1103/PhysRevLett.125.040501}
}

@Article{Paolo2019,
  author    = {Paolo, Agustin Di and Grimsmo, Arne L and Groszkowski, Peter and Koch, Jens and Blais, Alexandre},
  journal   = {New J. Phys.},
  title     = {{Control and coherence time enhancement of the 0-{$\pi$} qubit}},
  year      = {2019},
  issn      = {1367-2630},
  month     = apr,
  number    = {4},
  pages     = {043002},
  volume    = {21},
  doi       = {10.1088/1367-2630/ab09b0},
  publisher = {IOP Publishing},
}

@article{Rajabzadeh2023,
    title={{Analysis of arbitrary superconducting quantum circuits accompanied by a Python package: SQcircuit}},
    volume={7},
    ISSN={2521-327X},
    url={http://dx.doi.org/10.22331/q-2023-09-25-1118},
    journal={Quantum},
    publisher={Verein zur Forderung des Open Access Publizierens in den Quantenwissenschaften},
    author={Rajabzadeh, Taha and Wang, Zhaoyou and Lee, Nathan and Makihara, Takuma and Guo, Yudan and Safavi-Naeini, Amir H.},
    year={2023},
    month=sep, 
    pages={1118} 
}

@misc{Roverch2026,
  author        = {Roverc'h, Erwan and Borgognoni, Alvise and Villiers, Marius and Gerashchenko, Kyrylo and Smith, W. Clarke and Wilson, Christopher and Dou{\c{c}}ot, Benoit and Petrescu, Alexandru and Campagne-Ibarcq, Philippe and Leghtas, Zaki},
  title         = {{Experimental realization of a $\cos(2\varphi)$ transmon qubit}},
  year          = {2026},
  month         = mar,
  archiveprefix = {arXiv},
  copyright     = {Creative Commons Attribution 4.0 International},
  doi           = {10.48550/ARXIV.2603.13114},
  eprint        = {2603.13114},
  primaryclass  = {quant-ph},
  publisher     = {arXiv},
}

@article{Rymarz2021,
  title = {{Hardware-Encoding Grid States in a Nonreciprocal Superconducting Circuit}},
  author = {Rymarz, Martin and Bosco, Stefano and Ciani, Alessandro and DiVincenzo, David P.},
  journal = {Phys. Rev. X},
  volume = {11},
  issue = {1},
  pages = {011032},
  numpages = {25},
  year = {2021},
  month = {Feb},
  publisher = {American Physical Society},
  doi = {10.1103/PhysRevX.11.011032},
  url = {https://link.aps.org/doi/10.1103/PhysRevX.11.011032}
}

@article{Shagalov2026,
  title = {{Higher Josephson Harmonics in a Tunable Double-Junction Transmon Qubit}},
  author = {Shagalov, Ksenia and Feldstein-Bofill, David and Uhre Jakobsen, Leo and Sun, Zhenhai and Wied, Casper and Paulsen, Amalie T. J. and Bock Severin, Johann and Marciniak, Malthe A. and Potts, Clinton A. and Kringh{\o}j, Anders and Hastrup, Jacob and Flensberg, Karsten and Kr{\o}jer, Svend and Kjaergaard, Morten},
  journal = {Phys. Rev. Lett.},
  volume = {137},
  issue = {5},
  pages = {057001},
  numpages = {8},
  year = {2026},
  month = {Jul},
  publisher = {American Physical Society},
  doi = {10.1103/fr7x-555s},
  url = {https://link.aps.org/doi/10.1103/fr7x-555s}
}

@article{Shaw2024,
  title = {{Stabilizer Subsystem Decompositions for Single- and Multimode Gottesman-Kitaev-Preskill Codes}},
  author = {Shaw, Mackenzie H. and Doherty, Andrew C. and Grimsmo, Arne L.},
  journal = {PRX Quantum},
  volume = {5},
  issue = {1},
  pages = {010331},
  numpages = {36},
  year = {2024},
  month = {Feb},
  publisher = {American Physical Society},
  doi = {10.1103/PRXQuantum.5.010331},
  url = {https://link.aps.org/doi/10.1103/PRXQuantum.5.010331}
}

@Misc{Singh2025,
  author        = {Shraddha Singh and Baptiste Royer and Steven M. Girvin},
  title         = {{Towards Non-Abelian Quantum Signal Processing: Efficient Control of Hybrid Continuous- and Discrete-Variable Architectures}},
  year          = {2025},
  month         = apr,
  archiveprefix = {arXiv},
  eprint        = {2504.19992},
  primaryclass  = {quant-ph},
}

@article{Smith2020,
    title={{Superconducting circuit protected by two-Cooper-pair tunneling}},
    volume={6},
    ISSN={2056-6387},
    url={http://dx.doi.org/10.1038/s41534-019-0231-2},
    pages={8},
    journal={npj Quantum Inf.},
    publisher={Springer Science and Business Media LLC},
    author={Smith, W. C. and Kou, A. and Xiao, X. and Vool, U. and Devoret, M. H.},
    year={2020},
    month=jan 
}

@article{Smith2022,
  title = {{Magnifying Quantum Phase Fluctuations with Cooper-Pair Pairing}},
  author = {Smith, W. C. and Villiers, M. and Marquet, A. and Palomo, J. and Delbecq, M. R. and Kontos, T. and Campagne-Ibarcq, P. and Dou{\c{c}}ot, B. and Leghtas, Z.},
  journal = {Phys. Rev. X},
  volume = {12},
  issue = {2},
  pages = {021002},
  numpages = {11},
  year = {2022},
  month = {Apr},
  publisher = {American Physical Society},
  doi = {10.1103/PhysRevX.12.021002},
  url = {https://link.aps.org/doi/10.1103/PhysRevX.12.021002}
}

@article{Smith2025,
  author    = {Smith, W. C. and Borgognoni, A. and Villiers, M. and Roverc'h, E. and Palomo, J. and Delbecq, M. R. and Kontos, T. and Campagne-Ibarcq, P. and Dou{\c{c}}ot, B. and Leghtas, Z.},
  journal   = {Nat. Commun.},
  title     = {{Spectral signature of high-order photon processes enhanced by Cooper-pair pairing}},
  year      = {2025},
  issn      = {2041-1723},
  month     = sep,
  pages = {8359},
  number    = {1},
  volume    = {16},
  doi       = {10.1038/s41467-025-62047-8},
  publisher = {Springer Science and Business Media LLC},
}

@article{Touzard2019,
  author    = {Touzard, S. and Kou, A. and Frattini, N.E. and Sivak, V.V. and Puri, S. and Grimm, A. and Frunzio, L. and Shankar, S. and Devoret, M.H.},
  journal   = {Phys. Rev. Lett.},
  title     = {{Gated Conditional Displacement Readout of Superconducting Qubits}},
  year      = {2019},
  issn      = {1079-7114},
  month     = feb,
  number    = {8},
  pages     = {080502},
  volume    = {122},
  doi       = {10.1103/physrevlett.122.080502},
  publisher = {American Physical Society (APS)},
}

@article{Vool2017,
    title={{Introduction to quantum electromagnetic circuits}},
    volume={45},
    ISSN={1097-007X},
    url={http://dx.doi.org/10.1002/cta.2359},
    number={7},
    journal={Int. J. Circuit Theory Appl.},
    publisher={Wiley},
    author={Vool, Uri and Devoret, Michel},
    year={2017},
    month=jun, 
    pages={897} 
}

@Article{Vuillot2024,
  author    = {Vuillot, Christophe and Ciani, Alessandro and Terhal, Barbara M.},
  journal   = {Commun. Math. Phys.},
  title     = {{Homological Quantum Rotor Codes: Logical Qubits from Torsion}},
  year      = {2024},
  issn      = {1432-0916},
  month     = feb,
  number    = {2},
  volume    = {405},
  pages = {53},
  doi       = {10.1007/s00220-023-04905-4},
  publisher = {Springer Science and Business Media LLC},
}

@misc{Weissler2026,
    author = {Weissler, Eli J. and Smith, Thomas B. and Doherty, Andrew C. and Combes, Joshua},
    title = {{Structure-preserving transformations for superconducting circuits}},
    note = {in preparation},
}

@article{You2019,
   title={{Circuit quantization in the presence of time-dependent external flux}},
   volume={99},
   ISSN={2469-9969},
   url={http://dx.doi.org/10.1103/PhysRevB.99.174512},
   number={17},
   pages = {174512},
   numpages = {10},
   journal={Phys. Rev. B},
   publisher={American Physical Society (APS)},
   author={You, Xinyuan and Sauls, J. A. and Koch, Jens},
   year={2019},
   month=may 
}

@misc{ZenodoRepo,
    author = {Xanda C Kolesnikow and Thomas B Smith and Andrew C Doherty},
    title = {{Data and code for ``Protected measurements for protected superconducting qubits"}},
    journal = {Zenodo},
    year = {2026},
    url={https://doi.org/10.5281/zenodo.21729987}
}

@Book{Zwiebach2022,
  author    = {Zwiebach, Barton},
  publisher = {MIT Press},
  title     = {{Mastering Quantum Mechanics}},
  year      = {2022},
  isbn      = {9780262046138},
  pages     = {1200},
  subtitle  = {{Essentials, Theory, and Applications}},
  address = {Cambridge, MA, USA}
}
	
\end{document}